\documentclass[a4paper,11pt]{article}
\usepackage{jheppub} 
\usepackage{graphicx} % Required for inserting images
\usepackage{bbold}
\usepackage{subcaption}
\usepackage{amsmath}
\usepackage{gensymb}
\usepackage{booktabs}
\usepackage{pdflscape} 
\usepackage{rotating}
\usepackage{graphicx}
\usepackage{array}
\usepackage{makecell}

\newcommand{\rot}[1]{\rotatebox[origin=c]{90}{\scriptsize $#1$}}

\newcommand{\Op}[1]{\mathcal{O}^{#1}}

\title{Cartography of LNV dim-9 SMEFT: Implications for Radiative Neutrino Masses and $0\nu\beta\beta$}

\author[a]{Fabian Esser,}
\author[a,b]{Luk{\'a}{\v{s}} Gr{\'a}f,}
\author[c]{ and Chandan Hati} 

\affiliation[a]{Institute of Particle and Nuclear Physics (IPNP), Faculty of Mathematics and Physics, Charles University
Prague, V Holešovičkách 2, 180 00 Praha 8, Czech Republic}
\affiliation[b]{Institute of Physics, Silesian University in Opava, Bezru{\v{c}}ovo n{\'a}m{\v{e}}st{\'i} 1150/13, 746 01 Opava, Czech Republic}
\affiliation[c]{Instituto de F\'isica Corpuscular (IFIC), Universidad de Valencia-CSIC, E-46980 Valencia, Spain}

\emailAdd{fabian.esser@matfyz.cuni.cz}
\emailAdd{lukas.graf@matfyz.cuni.cz}
\emailAdd{chandan@ific.uv.es}

\abstract{
    We perform a systematic study of lepton-number–violating (LNV) dimension-9 operators in the Standard Model Effective Field Theory (SMEFT) that can mediate neutrinoless double beta decay ($0\nu\beta\beta$) at tree level, and map them to their possible tree-level ultraviolet completions. Using a diagram-based classification, we enumerate all such completions and isolate minimal two-particle models that avoid generating the dimension-5 Weinberg operator or dimension-7 LNV operators at tree level. We then chart how these minimal models populate the operator landscape and organise them by the loop order at which they radiatively induce lower-dimensional LNV operators, highlighting scenarios in which the tree-level dimension-9 contribution can compete with or dominate loop-suppressed neutrino-mass (dimension-5) effects. Representative one-loop and two-loop classes are matched onto the SMEFT, and their implications for neutrino masses, charged-lepton flavour violation, and the relative size of dimension-9 versus dimension-5 contributions to $0\nu\beta\beta$ are analysed, delineating regions of parameter space where upcoming experiments can be sensitive to genuinely short-range LNV dynamics.}

\begin{document}
\maketitle

\section{Introduction}
The conservation of lepton number is an accidental global symmetry of the Standard Model (SM) of particle physics. Observing processes that violate this symmetry would therefore constitute unambiguous evidence for physics beyond the Standard Model (BSM). Two well-known phenomena requiring violation of lepton number by two units are the generation of Majorana neutrino masses and neutrinoless double beta decay ($0\nu\beta\beta$)~\cite{Furry:1939qr}. The former can be most minimally achieved using the dimension-5 Weinberg operator, $\mathcal{O}_5 = \left(L^\dagger L\right) \, \left(H^\dagger H\right)$~\cite{Weinberg:1979sa}, which corresponds to the lowest odd dimension where lepton number violating can occur in Standard Model Effective Field Theory (SMEFT). The Weinberg operator can be realised at tree level via the three canonical seesaw mechanisms~\cite{Minkowski:1977sc,Gell-Mann:1979vob,Yanagida:1979as,Mohapatra:1979ia,Schechter:1980gr,Ma:1998dn}, but it may also arise radiatively at different loop orders~\cite{Bonnet:2012kz,AristizabalSierra:2014wal,Cepedello:2018rfh}. 

Neutrinoless double beta decay, $nn \to pp\, ee$, is analogous to ordinary double beta decay but proceeds without neutrinos in the final state. If the light (heavy) neutrinos are Majorana fermions, then the process can be mediated by the exchange of a (virtual) neutrino with a lepton-number–violating (LNV) Majorana mass insertion between two weak-interaction vertices (following a mass-mixing). \footnote{Additional new light physics can even lead to other, alternative modes of this LNV process~\cite{Doi:1987rx,Graf:2023dzf,deVries:2025hqa}.} However, within the Standard Model Effective Field Theory (SMEFT) $0\nu\beta\beta$ can also be induced at tree level by other higher-dimensional lepton-number–violating (LNV) contact or long-range operators \cite{Pas:1999fc,Pas:2000vn,Deppisch:2012nb,Bonnet:2012kh,Deppisch:2017ecm,Cirigliano:2017ymo,Cirigliano:2017djv,Graf:2018ozy,Cirigliano:2018hja,Cirigliano:2018yza,Deppisch:2020ztt}. For a recent systematic study of dimension-7 LNV SMEFT operators in the context of $0\nu\beta\beta$ and other LNV observables across different energy scales, see for instance~Ref.~\cite{Fridell:2023rtr}. Importantly, realistic ultraviolet (UV) completions typically generate not just a single operator but correlated sets of operators. For example, higher-dimensional operators may appear at tree level, while the Weinberg operator is generated only radiatively. An analysis of UV completions triggering dimension-7 operators at tree level, along with loop-level realisation of dimension-5 operators, can be found in Ref.~\cite{Fridell:2024pmw}. 

In the present work, we systematically explore the tree-level UV realisations of LNV dimension-9 SMEFT operators in the context of radiative neutrino masses, identifying the interesting scenarios in which the dimension-5 and dimension-7 LNV SMEFT operators are generated only at the loop level. We also study in detail some model classes in the context of $0\nu \beta \beta$, neutrino oscillation data, and charged lepton flavour observables.

This article is organised as follows: In Sec.~\ref{sec:smeft} we briefly summarise the EFT framework and list the LNV SMEFT operators at dimension-9, dimension-7, and dimension-5.
In Sec.~\ref{sec:classification}, we present a diagram-based method to open up any SMEFT operator, and use this method to list and classify all tree-level models for the LNV dimension-9 operators. Excluding models that generate dimension-5 or dimension-7 operators at tree level, we identify minimal models and classify them according to the operators they generate in Sec.~\ref{sec:minimal_models}. Next, in Sec.~\ref{sec:1L_model} we analyse the class of minimal models that generate the Weinberg operator at 1-loop level, and calculate the matching to the SMEFT for the dimension-5, dimension-7 and dimension-9 operators and the induced $0\nu\beta\beta$ by each operator dimension for a representative example model. In Sec.~\ref{sec:2L_model}, we discuss minimal models that generate the Weinberg operator at 2-loop level, present the matching for a representative example model and use a Casas-Ibarra parametrisation to explain the observed neutrino masses and mixing. For different flavour assumptions of the couplings, we  evaluate the constraints from charged lepton flavour violation (cLFV) and the relative contributions of the various induced operators to $0\nu\beta\beta$. This allows us to delineate regions of the parameter space in which the dimension-9 contributions dominate over the loop-suppressed dimension-5 terms. We conclude in Sec.~\ref{sec:conclusion}.

\newpage
\section{Lepton number violation in the SMEFT}
\label{sec:smeft}

\begin{table}[b]
\centering
\renewcommand{\arraystretch}{0.85}
\setlength{\tabcolsep}{6pt}

\begin{tabular}{llc}
Class & Name & Operator \\
\midrule
$\Psi^2H^2$
& $\Op{(5)pr}_{LH}$
& $\epsilon_{ik} \epsilon_{jl}\, \overline{L_{i,p}^C} L_{j,r} H_k H_l$ \\
\bottomrule
\end{tabular}
\caption{Dimension-5 operators with $\Delta L = 2$: the Weinberg operator.}
\label{tab:operators_dim5}

\renewcommand{\arraystretch}{1.0}

\end{table}

\begin{table}[t]
\centering

\renewcommand{\arraystretch}{0.9}
\setlength{\tabcolsep}{6pt}

\begin{tabular}{llc}
Class & Name & Operator \\
\midrule

$\Psi^2H^4$
& $\Op{(7)pr}_{LH}$
& $\epsilon_{ik}\epsilon_{jl}\,\overline{L_{i,p}^C}\,L_{j,r}\,H_k H_l (H^\dagger H)$
\\[4pt]
\cmidrule(l){2-3}

$\Psi^2H^3 D$
& $\Op{(7)pr}_{LHDe}$
& $\epsilon_{ik}\epsilon_{jl}\,\overline{L_{i,p}^C}\gamma^\mu e_r\,H_j H_k (D_\mu H_l)$
\\[4pt]
\cmidrule(l){2-3}

$\Psi^2H^2D^2$
& $\Op{(7)pr}_{LHD1}$
& $\epsilon_{ij}\epsilon_{kl}\,\overline{L_{i,p}^C}(D_\mu L_{j,r})\,H_k (D^\mu H_l)$
\\
& $\Op{(7)pr}_{LHD2}$
& $\epsilon_{ik}\epsilon_{jl}\,\overline{L_{i,p}^C}(D_\mu L_{j,r})\,H_k (D^\mu H_l)$
\\[4pt]
\cmidrule(l){2-3}

$\Psi^2H^2W$
& $\Op{(7)pr}_{LHW}$
& $\epsilon_{ik}\epsilon_{jl}\,g_2\,\tau^I_{ml}\,\overline{L_{i,p}^C}\sigma^{\mu\nu}L_{j,r}\,H_k H_l\,W_{\mu\nu}^I$
\\[4pt]
\cmidrule(l){2-3}

$\Psi^4 D$
& $\Op{(7)prst}_{LLduD1}$
& $\epsilon_{ij}\,\overline{d_{a,s}}\gamma^\mu u_{a,t}\,\overline{L_{i,p}^C}(D_\mu L_{j,r})$
\\[4pt]
\cmidrule(l){2-3}

$\Psi^4 H$
& $\Op{(7)prst}_{LLQdH1}$
& $\epsilon_{ik}\epsilon_{jl}\,\overline{d_{a,t}}L_{i,p}\,\overline{Q_{ja,s}^C}L_{k,r}\,H_l$
\\
& $\Op{(7)prst}_{LLQdH2}$
& $\epsilon_{ij}\epsilon_{kl}\,\overline{d_{a,t}}L_{i,p}\,\overline{Q_{ja,s}^C}L_{k,r}\,H_l$
\\
& $\Op{(7)prst}_{LLQuH}$
& $\epsilon_{ij}\,\overline{Q_{ka,s}} u_{a,t}\,\overline{L_{k,p}^C}L_{i,r}\,H_j$
\\
& $\Op{(7)prst}_{LeudH}$
& $\epsilon_{ij}\,\overline{L_{i,p}^C}\gamma^\mu e_r\,\overline{d_{a,t}}\gamma_\mu u_{a,s}\,H_j$
\\

\bottomrule
\end{tabular}
\caption{Dimension-7 operators with $\Delta L =2$ that are generated at 1-loop level by the example models discussed. Here, $\{i,j,k,l\}$, $\{a\}$ and $\{p,r,s,t\}$ denote $SU(2)$, colour and flavour indices, respectively.}
\label{tab:operators_dim7}

\renewcommand{\arraystretch}{1.0}

\end{table}

\begin{table}[t]
\centering

\renewcommand{\arraystretch}{0.85}
\setlength{\tabcolsep}{6pt}

\begin{tabular}{llc}
Class & Name & Operator \\
\midrule
$\Psi^6$ 
& $\Op{(9)prstuv}_{ddueue}$ 
& $\overline{d_{a,p}} d_{b,r}^{C}\, \overline{u_{a,s}^{C}} e_{t}\, \overline{u_{b,u}^{C}} e_{v}$ \\
\cmidrule(l){2-3}

& $\Op{(9)prstuv}_{dQdueL1}$ 
& $\epsilon_{ij}\, \overline{d_{a,p}} Q_{ia,r}\, \overline{d_{b,s}} \gamma_{\mu} u_{b,t}\, \overline{e_{u}^{C}} \gamma^{\mu} L_{j,v}$ \\
& $\Op{(9)prstuv}_{dQdueL2}$ 
& $\epsilon_{ij}\, \overline{d_{a,p}} Q_{ib,r}\, \overline{d_{b,s}} \gamma_{\mu} u_{a,t}\, \overline{e_{u}^{C}} \gamma^{\mu} L_{j,v}$ \\
\cmidrule(l){2-3}

& $\Op{(9)prstuv}_{QudueL1}$ 
& $\overline{Q_{ia,p}} u_{a,r}\, \overline{d_{b,s}} \gamma_{\mu} u_{b,t}\, \overline{e_{u}^{C}} \gamma^{\mu} L_{i,v}$ \\
& $\Op{(9)prstuv}_{QudueL2}$ 
& $\overline{Q_{ia,p}} u_{b,r}\, \overline{d_{b,s}} \gamma_{\mu} u_{a,t}\, \overline{e_{u}^{C}} \gamma^{\mu} L_{i,v}$ \\
\cmidrule(l){2-3}

& $\Op{(9)prstuv}_{dQdQLL1}$ 
& $\epsilon_{ik}\epsilon_{jl}\, \overline{d_{a,p}} Q_{ia,r}\, \overline{d_{b,s}} Q_{jb,t}\, \overline{L_{k,u}^{C}} L_{l,v}$ \\
& $\Op{(9)prstuv}_{dQdQLL2}$ 
& $\epsilon_{ik}\epsilon_{jl}\, \overline{d_{a,p}} Q_{ib,r}\, \overline{d_{b,s}} Q_{ja,t}\, \overline{L_{k,u}^{C}} L_{l,v}$ \\
\cmidrule(l){2-3}

& $\Op{(9)prstuv}_{dQQuLL1}$ 
& $\epsilon_{ij}\, \overline{d_{a,p}} Q_{ia,r}\, \overline{Q_{kb,s}} u_{b,t}\, \overline{L_{k,u}^{C}} L_{j,v}$ \\
& $\Op{(9)prstuv}_{dQQuLL2}$ 
& $\epsilon_{ij}\, \overline{d_{a,p}} Q_{ib,r}\, \overline{Q_{kb,s}} u_{a,t}\, \overline{L_{k,u}^{C}} L_{j,v}$ \\
\cmidrule(l){2-3}

& $\Op{(9)prstuv}_{QuQuLL1}$ 
& $\overline{Q_{ia,p}} u_{a,r}\, \overline{Q_{jb,s}} u_{b,t}\, \overline{L_{i,u}^{C}} L_{j,v}$ \\
& $\Op{(9)prstuv}_{QuQuLL2}$ 
& $\overline{Q_{ia,p}} u_{b,r}\, \overline{Q_{jb,s}} u_{a,t}\, \overline{L_{i,u}^{C}} L_{j,v}$ \\
\midrule

$\Psi^2H^4W$ 
& $\Op{(9)p r}_{LLH^4W1}$ 
& $\epsilon_{ik} \epsilon_{jl} g_2 \tau^I_{ml}\, \overline{L_{i,p}^C} \sigma^{\mu\nu} L_{jr}\, H_k H_l W_{\mu\nu}^I (H^{\dagger} H)$ \\
\midrule

$\Psi^2H^4D^2$ 
&  $\Op{(9)p r}_{eeH^4D^2}$ 
& $\epsilon_{ij} \epsilon_{kl}\, \overline{e_p^C} e_r\, H_i D_{\mu} H_j\, H_k D^{\mu} H_l$   \\ 

& $\Op{(9)p r}_{LLH^4D^23}$ 
& $\epsilon_{ik} \epsilon_{jl}\, \overline{D_{\mu} L_{i,p}^C}\, D_{\mu} L_{j,r}\, H_k H_l (H^{\dagger} H)$ \\

&  $\Op{(9)p r}_{LLH^4D^24}$ 
& $\epsilon_{ik} \epsilon_{jl}\, \overline{L_{i,p}^C}\, D_{\mu} L_{j,r}\, D_{\mu} H_k\, H_l (H^{\dagger} H)$ \\
\midrule

$\Psi^4 H^2 D$ 
& $\Op{(9)prst}_{deueH^2D}$ 
& $\epsilon_{ij}\, \overline{d_{a,p}} \gamma^{\mu} e_r\, \overline{u_{a,s}^C} e_t\, H_i i D_{\mu} H_j$ \\ 
\cmidrule(l){2-3}

&  $\Op{(9)prst}_{dQLeH^2D2}$ 
& $\epsilon_{ik} \epsilon_{jl}\, \overline{d_{a,p}} Q_{ia,r}\, \overline{L_{j,s}^C} \gamma^{\mu} e_t\, H_k i D_{\mu} H_l$ \\

&  $\Op{(9)prst}_{dLQeH^2D1}$ 
& $\epsilon_{ik} \epsilon_{jl}\, \overline{d_{a,p}} L_{i,r}\, \overline{Q_{ja,s}^C} \gamma^{\mu} e_t\, i D_{\mu} H_k\, H_l$ \\
\cmidrule(l){2-3}

&  $\Op{(9)prst}_{dLuLH^2D2}$ 
& $\epsilon_{ik} \epsilon_{jl}\, \overline{d_{a,p}} L_{i,r}\, \overline{u_{a,s}^C} \gamma^{\mu} L_{j,t}\, i D_{\mu} \tilde{H}_k\, H_l$ \\

&  $\Op{(9)prst}_{duLLH^2D}$ 
& $\epsilon_{ik} \epsilon_{jl}\, \overline{d_{a,p}} \gamma^{\mu} u_{a,r}\, \overline{L_{i,s}^C} i D_{\mu} L_{j,t}\, \tilde{H}_k H_l$ \\
\cmidrule(l){2-3}

&  $\Op{(9)prst}_{deQLLH^2D}$ 
& $\epsilon_{ik} \epsilon_{jl}\, \overline{d_{a,p}} \gamma^{\mu} e_r\, \overline{Q_{ia,s}^C} i D_{\mu} L_{j,t}\, H_k H_l$ \\
\cmidrule(l){2-3}

&  $\Op{(9)prst}_{QueLH^2D2}$ 
& $\epsilon_{jk}\, \overline{Q_{ia,p}} u_{a,r}\, \overline{e_{s}^C} \gamma^{\mu} L_{j,t}\, H_i i D_{\mu} H_k$ \\

&  $\Op{(9)prst}_{QeuLH^2D2}$ 
& $\epsilon_{jk}\, \overline{Q_{ia,p}} e_{r}\, \overline{u_{a,s}^C} \gamma^{\mu} L_{j,t}\, H_i i D_{\mu} H_k$ \\
\cmidrule(l){2-3}

&  $\Op{(9)prst}_{QLQLH^2D2}$ 
& $\epsilon_{jl} \epsilon_{km}\, \overline{Q_{ia,p}} \gamma^{\mu} L_{i,r}\, \overline{Q_{ja,s}^C} \gamma_{\mu} L_{k,t}\, H_l i D_{\mu} H_m$ \\

&  $\Op{(9)prst}_{QLQLH^2D5}$ 
& $\epsilon_{jl} \epsilon_{km}\, \overline{Q_{ia,p}} \gamma^{\mu} L_{j,r}\, \overline{Q_{ka,s}^C} \gamma_{\mu} L_{i,t}\, i D_{\mu} H_l\, H_m$ \\

&  $\Op{(9)prst}_{QQLLH^2D2}$ 
& $\epsilon_{jl} \epsilon_{km}\, \overline{Q_{ia,p}} \gamma^{\mu} Q_{ja,r}\, \overline{L_{i,s}^C} i D_{\mu} L_{k,t}\, H_l H_m$ \\
\bottomrule
\end{tabular}
\caption{Dimension-9 operators with $\Delta L =2$ that are generated at tree-level by the example models discussed. Here, $\{i,j,k,l\}$, $\{a,b\}$ and $\{p,r,s,t,u,v\}$ denote $SU(2)$, colour and flavour indices, respectively.}
\label{tab:operators_dim9}

\renewcommand{\arraystretch}{1.0}

\end{table}

Effective Field Theories (EFTs) provide a systematic framework to describe and compute a process at a given energy scale ($E$) in terms of only the degrees of freedom relevant at those energies with masses $m \lesssim E$, relying on the decoupling of heavy new degrees of freedom with masses $M\sim\Lambda \gg E$. At the energy $E$, only the light degrees of freedom with masses $m \lesssim E$ are considered as physical degrees of freedom, and the heavy degrees of freedom are ``integrated out". Such an effective theory can be parametrised in terms of a series of local interactions among only the light particles, $\mathcal{L}_{EFT} = \sum_i c_i \mathcal{O}_i$, where only the Wilson coefficients $c_i \propto \frac{1}{\Lambda^{d_i-4}}$ encode the details of the heavy physics.

In the \textit{top-down} EFT approach, where we know the full UV theory, the Wilson coefficients $c_i$ in the EFT are determined by \textit{matching} the full theory to the EFT at the scale $\mu=\Lambda\sim M$. Since physical observables are built out of correlation functions, it is crucial to ensure that the one-light-particle irreducible diagrams/action computed from $\mathcal{L}_{\text{EFT}}$ match with $\mathcal{L}_{\text{UV}}$ at $\mu=M$, known as the ``matching condition". In practice, there are two popular techniques for performing the matching: \textit{diagrammatic} and \textit{functional}. In case of diagrammatic matching, one calculates all irreducible Feynman diagrams for the relevant processes in both the full theory and the EFT and matches them at $\mu=\Lambda\sim M$~\cite{Pich:1998xt}. On the other hand, in the functional matching approach, one evaluates the path integral by expanding the contributions of the heavy fields to the functional determinant to calculate the effective action in both theories and matches them at $\mu=\Lambda\sim M$~\cite{Henning:2014wua, Gaillard:1985uh,CHEYETTE1988183}. The diagrammatic approach is used by the automated package \texttt{MatchMakerEFT}~\cite{Carmona:2021xtq}, which is currently automated up to dimension-6 for the SMEFT, and the functional approach is employed by the automated package \texttt{Matchete}~\cite{Fuentes-Martin:2022jrf}, which provides up to one-loop matching for any arbitrary dimension in an arbitrary Green's basis.

After obtaining the matching of the UV model to a given EFT basis at a scale $\mu=M\equiv \Lambda$, as described above, one further needs to evolve the Wilson coefficients down to the scale of the experimental observable. This evolution is accounted for by the renormalisation group equation (RGE) running, which can also induce mixing between operators. 

The SMEFT is an EFT valid above the electroweak symmetry breaking scale, respecting the whole unbroken SM gauge group $SU(3)_C\otimes SU(2)_L\otimes U(1)_Y$ and consists of only the Standard Model (SM) field content as its degrees of freedom. The generic Lagrangian for the SMEFT can be expressed as
\begin{equation}
\mathcal{L}_{SMEFT} = \mathcal{L}_{SM} +   
c_5 \mathcal{O}_5 + \sum_{i} c_{6i} \mathcal{O}_{6i} + \sum_{i} c_{7i} \mathcal{O}_{7i} + \sum_{i}  
c_{8i} \mathcal{O}_{8i} + \sum_{i}  
c_{9i} \mathcal{O}_{9i} + \ldots,
\end{equation}
where $\mathcal{O}_d$ represents $d$-dimensional nonrenormalisable operators constructed out of the SM fields and the corresponding Wilson coefficient $c_d \propto \Lambda^{4-d}$ encodes the effects of the heavy new physics degrees of freedom\footnote{The SMEFT framework can be conveniently used for model-independent interpretations of experimental data from searches for new physics. The Wilson coefficients can be fitted to the experimental data, directly or via a chain of low-energy EFTs, independently of any assumptions on the BSM model. This can be achieved by performing global SMEFT fits, cf.~e.g.~\cite{Ellis:2020unq, Giani:2023gfq} or by likelihood analyses for specific observables.}.

The operators that are relevant for Majorana neutrino masses and $0\nu\beta\beta$ violate lepton number by two units, $\Delta L = 2$, and appear only at odd dimensions in the SMEFT~\cite{Kobach:2016ami}. Tables~\ref{tab:operators_dim5}, \ref{tab:operators_dim7} and \ref{tab:operators_dim9} show the dimension-5, dimension-7 and dimension-9 operators that contribute at tree level to $0\nu\beta\beta$~\cite{Scholer:2023bnn}. We denote the colour, the $SU(2)$ and the flavour indices by $\{a, b\}$, $\{i,j,k,l,m\}$ and $\{p,r,s,t,u,v\}$, respectively. In this work, we systematically explore the dimension-9 LNV SMEFT operators to identify tree-level UV completions that can explain neutrino masses and lead to observable $0\nu\beta\beta$. In particular, we are interested in UV models that induce the Weinberg operator at higher-loop order while triggering dimension-9 operators at tree level.  UV completions for LNV dimension-7 operators have been discussed e.g.~in~\cite{Fridell:2023rtr, Fridell:2024pmw, Graf:2025cfk}. In particular, in~\cite{Graf:2025cfk} it was recently shown that the dimension-7 operators mix with each other and with the Weinberg operator via renormalisation group equation (RGE) running, and that the underlying UV models can induce enhanced matching contributions to the Weinberg operator at the loop level, thus leading to stronger individual limits on the Wilson coefficients. In this work, we focus on matching effects and study models inducing the $0\nu\beta\beta$-triggering dimension-9 operators at tree level, see Tab.~\ref{tab:operators_dim9}. One of the key objectives of this work is to identify the potentially interesting scenarios where the scale suppression of the dimension-9 operators competes with the loop suppression of the Weinberg operator, and to study such classes of models to determine the relative dominance of dimension-9 tree-level $0\nu\beta\beta$ contributions over loop-induced dimension-5 contributions also in context of neutrino oscillation data and charged lepton flavour violating observables (cLFV).

\section{Diagram-based approach to classify UV completions for LNV SMEFT operators}
\label{sec:classification}

\subsection{Diagram based approach}
We are interested in a systematic method to find all possible tree-level UV completions for dimension-9 LNV SMEFT operators and, at the same time, in the minimal loop order at which such UV completions will induce dimension-5 and dimension-7 LNV operators. Although it is fairly straightforward to work out the tree-level UV completions for dimension-9 LNV SMEFT operators, it becomes quite nontrivial to systematically work out the minimal loop order for dimension-5 and dimension-7 LNV operators and to classify all possibilities. To this end, 
the diagram based approach that we follow aims to find UV completions by working out the Feynman diagrams systematically and consists of three main steps: 
\begin{enumerate}
    \item[(i)] \textbf{Topologies:} For a given operator, find all topologies with $k$ loops and $n$ external legs.
    \item[(ii)] \textbf{Diagrams:} Insert fermions, scalars and vectors in all possible ways allowed by Lorentz invariance into the topologies and keep only renormalisable interactions.
    \item[(ii)] \textbf{Model diagrams:} Insert all possible representations for fermions, scalars and vectors into the diagrams, i.e., assign the particles' quantum numbers.
\end{enumerate}

\begin{figure}[h!]
    \centering
\includegraphics[scale=0.5]{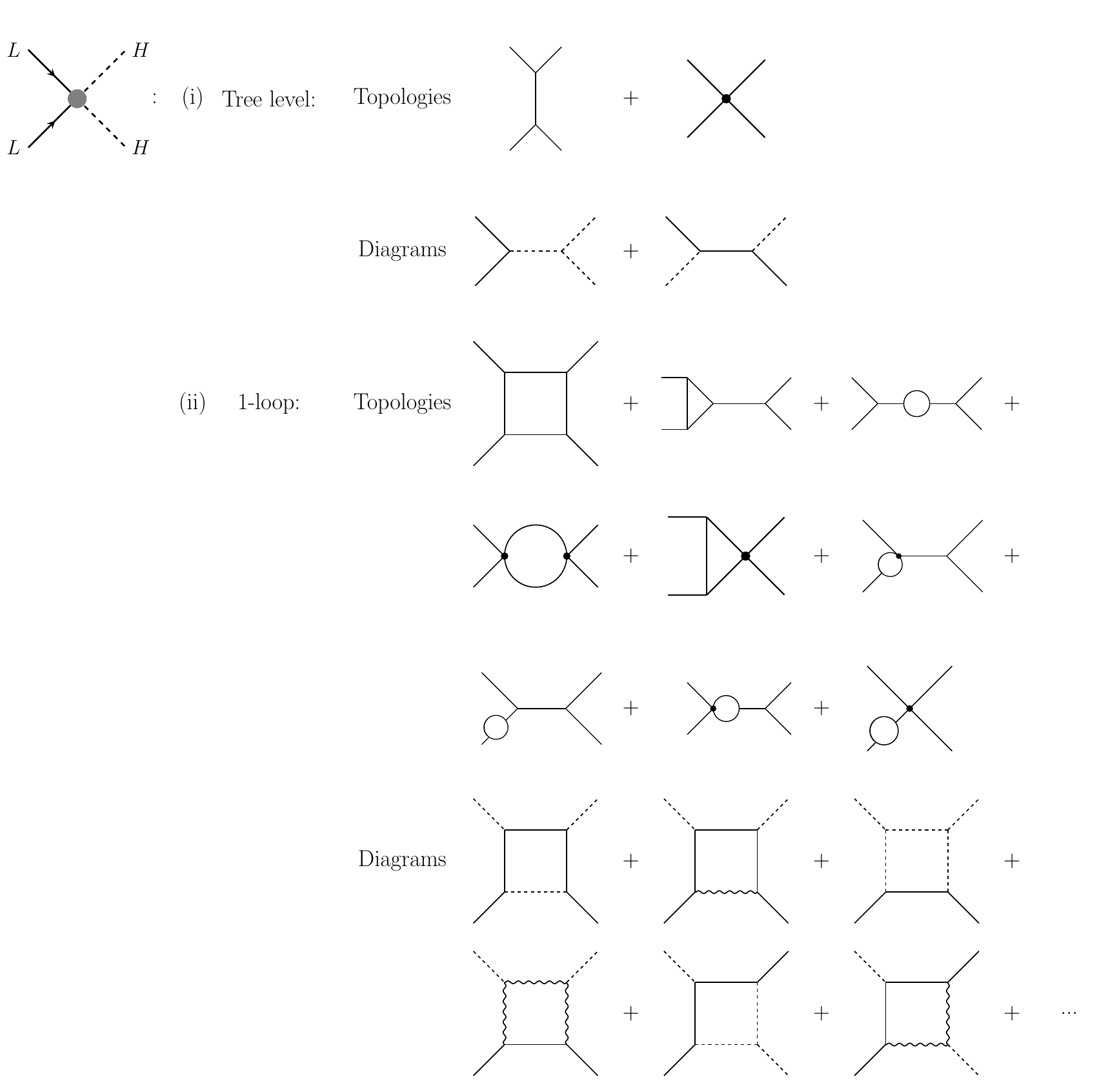}
    \caption{Step (i) and (ii) of the diagram based approach: topologies and diagrams at tree level and 1-loop level for the Weinberg operator.}    
\label{fig:topologies}
\end{figure}

In the first step, one finds, for each given operator, all topologies with the correct number of external legs at the specific loop-order we are interested in. For example, for the Weinberg operator with 4 external legs, the possible tree-level and 1-loop-level topologies are shown in~Fig.~\ref{fig:topologies}.
In the second step, one creates diagrams by inserting fermions, scalars and vectors into the topologies in all possible ways allowed by Lorentz invariance. For the full theory to be renormalisable, we keep only diagrams in which all vertices have a mass dimension of at most 4. For the Weinberg operator, we insert 2 scalars and 2 fermions into the external legs, leading to the diagrams shown in Fig.~\ref{fig:topologies}, where we denote scalars, fermions and vectors by dashed, solid and wavy lines, respectively. The second tree-level topology in the first row would, in this case, contain a dimension-5 vertex and is hence discarded, as well as the first 1-loop topology in the second row and the third in the third row. At the one-loop level, we show only the box diagrams emerging from the first topology for illustration purposes. 

\begin{figure}[h!]
    \centering
\includegraphics[scale=0.63]{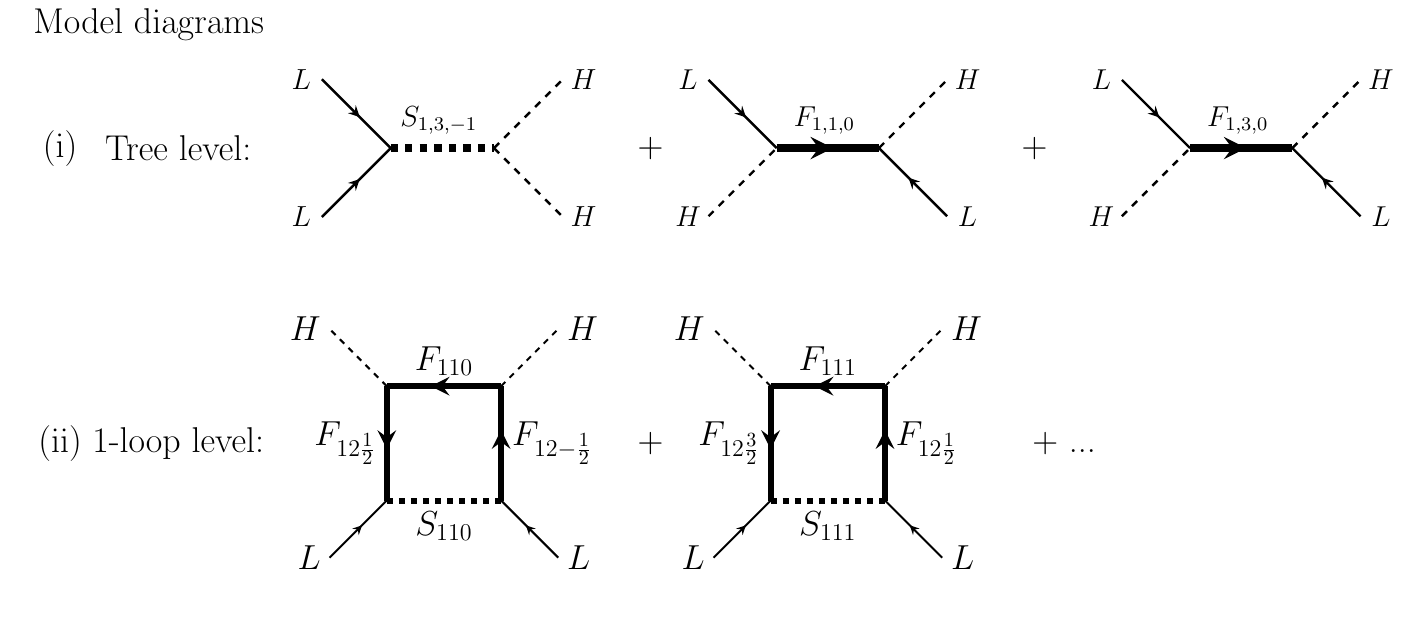}
    \caption{Step (iii) of the diagram based approach: model diagrams at tree level and 1-loop level for the Weinberg operator.}    
\label{fig:model_diagrams}
\end{figure}

Finally, in the third step, model diagrams are obtained by inserting specific representations for all quantum numbers into the diagrams. We use the notation $X_{a,b,c}$ to denote scalars (S), fermions (F) and vectors (V) with colour charge $a$, $SU(2)$ multiplicity $b$ and hypercharge $c$. Fig.~\ref{fig:model_diagrams} shows the model diagrams for the Weinberg operator. In the tree-level case, we recover the three well-known seesaw models with $F_{1,1,1}$ (type-I), $S_{1,3,-1}$ (type-II) and $F_{1,3,0}$ (type-III). At 1-loop, we obtain in principle infinitely many model diagrams, because the external particles only fix the colour, $SU(2)$ multiplicity and hypercharge quantum numbers up to a global seed in the loop. For example, the external Higgs and lepton doublets for the Weinberg operator determine the differences between hypercharges, and both model diagrams shown in Fig.~\ref{fig:model_diagrams} are valid examples for UV models. In practice, one restricts the number of models by imposing maximal representations $(\text{max}_{SU(3)}, \text{max}_{SU(2)}, \text{max}_{Y})$ for colour, $SU(2)$ and hypercharge, respectively.

One consequence of finding UV completions via Feynman diagrams is that different operators with the same external legs yield the same diagrams. This implies that this method completes operator structures with the same external legs rather than the operators themselves, and does not distinguish between different internal $SU(2)$ or colour contractions. When searching for models for e.g.~$\mathcal{O}_{dQdueL1}^{(9)}$ we search for topologies, diagrams and model diagrams with the external particles $d$, $Q$, $d$, $u$, $e$ and $L$. Whether this model diagram actually produces $\mathcal{O}_{dQdueL1}^{(9)}$, $\mathcal{O}_{dQdueL2}^{(9)}$, or both operators, has to be determined later by an explicit matching calculation. Hence, we group all operators with the same external particle content together and search for models for these \textit{operator structures} instead of individual operators. 

Another caveat appears for operators with derivatives. While the operator basis is defined in the unbroken phase of the SM and keeps the covariant derivative $D_{\mu}$ as a fundamental building block, the diagrams are built with either $\partial_{\mu}$ or the explicit gauge fields $G_{\mu}^a$, $W_{\mu}^I$ and $B_{\mu}$. This means that a term $\partial_{\mu} W_{\nu}$ can originate from both $\mathcal{O}_{LLH^4W1}^{(9)}$ and $\mathcal{O}_{LLH^4D^2}^{(9)}$. Therefore, we group these two operators together as a single operator structure $\mathcal{O}_{LLH^4(D/W)^2}^{(9)}$. Similarly, $\mathcal{O}_{LH(D/W)}^{(7)}$ contains both $\mathcal{O}_{LHD}^{(7)}$ and $\mathcal{O}_{LHW}^{(7)}$. Table~\ref{tab:operator_structures} lists all the operator structures for which we will classify the UV models. There are 14 structures at dimension 9, 7 at dimension 7 and 1 (Weinberg operator) at dimension 5.

\begin{table}[t]
\centering
\begin{tabular}{ll}
\toprule
\multicolumn{2}{l}{\textbf{Dimension 9}} \\
\midrule
$\Psi^6$ 
& $\mathcal{O}_{ddueue}^{(9)}$, $\mathcal{O}_{dQdueL}^{(9)}$, $\mathcal{O}_{QudueL}^{(9)}$, \\
& $\mathcal{O}_{dQdQLL}^{(9)}$, $\mathcal{O}_{dQQuLL}^{(9)}$, $\mathcal{O}_{QuQuLL}^{(9)}$ \\[4pt]

$\Psi^4 H^2 D$ 
& $\mathcal{O}_{deueH^2D}^{(9)}$, 
$\mathcal{O}_{dQLeH^2D}^{(9)}$,
$\mathcal{O}_{duLLH^2D}^{(9)}$, \\
& $\mathcal{O}_{deQLH^2D}^{(9)}$, $\mathcal{O}_{QueLH^2D}^{(9)}$, $\mathcal{O}_{QQLLH^2D}^{(9)}$  \\[4pt]

$\Psi^2 H^4 (D/W)^2$ 
& $\mathcal{O}_{eeH^4D^2}^{(9)}$, $\mathcal{O}_{LLH^4(D/W)^2}^{(9)}$  \\
\midrule
\multicolumn{2}{l}{\textbf{Dimension 7}} \\
\midrule
$\Psi^2 H^4$ 
& $\mathcal{O}_{LH}^{(7)}$ \\ 

$\Psi^2 H^3 D$ 
& $\mathcal{O}_{LHDe}^{(7)}$ \\

$\Psi^4 D$ 
& $\mathcal{O}_{LLduD}^{(7)}$ \\

$\Psi^4 H$ 
& $\mathcal{O}_{LLQdH}^{(7)}$,
$\mathcal{O}_{LLQuH}^{(7)}$,
$\mathcal{O}_{LeudH}^{(7)}$ \\
$\Psi^2 H^2 (D/W)^2$ 
& $\mathcal{O}_{LH(D/W)}^{(7)}$ \\
\midrule
\multicolumn{2}{l}{\textbf{Dimension 5}} \\
\midrule
$\Psi^2 H^2$ 
& $\mathcal{O}_{LH}^{(5)}$ \\
\bottomrule
\end{tabular}
\caption{Operator structures for $0\nu\beta\beta$ at dimension 9, 7 and 5.}
\label{tab:operator_structures}
\end{table}

At a given loop level $k\ge 1$, we call a model diagram \textit{non-genuine} if it corresponds to a diagram at loop level $k-1$ in which one vertex is generated at 1-loop level \cite{Bonnet_2012, AristizabalSierra:2014wal}.
\textit{Genuine} diagrams do not lead to such an effective contribution at a lower loop level. In the following classification, we focus on genuine model diagrams.

The diagram-based method described here can, in principle, be used by hand, but beyond tree level and for operators with many external legs, automation is advantageous. We use the prescription developed for the Weinberg operator in~\cite{Cepedello:2018rfh} and further extended for other classes of SMEFT operators, in particular for 4-fermion operators~\cite{Cepedello:2022pyx, Cepedello:2023yao} and neutral triple gauge couplings~\cite{Cepedello:2024ogz, Cepedello:2024qmq}. The relevant code, based on \texttt{GroupMath}, represents diagrams via generalised adjacency matrices and, in principle, works up to any loop order, though it becomes computationally more demanding with increasing loop order.

\subsection{Counting model diagrams and models for dimension-9 operators}

Now we are all set to apply the diagram-based approach presented in the previous subsection to find all models for the dimension-9 operator structures in Tab.~\ref{tab:operator_structures}, i.e.~the dimension-9 operators that can mediate $0\nu\beta\beta$ at tree level. By \textit{model} we denote the set of all BSM particles in a model diagram. Note that at the tree level, the number of models is finite, since all internal quantum numbers are fixed by those of the external fields. We find that all UV completions contain at most 4 different BSM particles and that for the operator class $\Psi^6$, cf.~Tab.~\ref{tab:operator_structures}, 3 different BSM particles are always sufficient. In the following, we will focus on the models with only 2 distinct BSM particles, which we will refer to as \textit{minimal models}, since we did not find any 1-particle BSM extension that can generate the dimension-9 operators. As shown in the first row of Tab.~\ref{tab:counting_models}, we find a total of 1474 models, i.e., distinct UV completions, of which 1222 contain at most 3 BSM particles, and 83 contain 2 BSM particles. 

In the next step, we scan over these models and exclude all models which generate any of the dimension-7 or dimension-5 operators in Tab.~\ref{tab:operator_structures} at tree-level. The rationale is that if any model generates a dimension-5 or dimension-7 operator at tree level, we expect the lower-dimensional contributions to $0\nu\beta\beta$ to be dominant. To exclude these models, we first generate all tree-level completions $\mathbb{T}_{7+5}$ for the dimension-7 operators and the Weinberg operator (i.e.~the seesaw models).
We then keep a model $A \in \mathbb{T}_9$, the tree-level models for the dimension-9 operators, if there is no subset of particles $B \subseteq A$ such that $B \in \mathbb{T}_{7+5}$. As shown in the second row of Tab.~\ref{tab:counting_models}, this restriction leaves us with 315 models, of which 301 incorporate 3 BSM particles and 29 contain 2 BSM particles. 

\begin{table}[t]
\centering
\begin{tabular}{lccc}
\toprule
 & $\leq$ 4P & $\leq$ 3P & 2P \\
\midrule
All tree-level dimension-9 models 
& 1474 & 1222 & 83 \\
No dimension-5 or dimension-7 at tree-level 
& 315 & 301 & 29 \\
No dimension-5 or dimension-7 at tree-level, no vectors 
& 75 & 75 & \textbf{14} \\
\bottomrule
\end{tabular}
\caption{Numbers of models (i.e.~different UV completions) for dimension-9 operators with $\Delta L = 2$ that can induce $0\nu\beta\beta$.}
\label{tab:counting_models}
\end{table}

In the final step, we restrict ourselves to models with BSM scalars and fermions, but no vectors. Vectors are more involved, because in a consistent UV completion they should appear from the breaking of a local gauge symmetry, i.e.~live in the adjoint representation of a gauge group. Firstly, some of the minimal gauge groups for given quantum numbers, cf.~e.g.~\cite{Fonseca_2017}, are quite exotic and nontrivial to embed in well-motivated grand unified groups.
Secondly, for dimension-9 operators to be observable at experiments, like searches for $0\nu\beta\beta$, the relevant new physics scale should be around TeV, in contrast to the typical grand unification scales allowed by the usual proton decay constraints. Therefore, realisations of vector UV completions require the implementation of more involved, non-minimal intermediate symmetry-breaking chains and/or flavour structures.

As listed in the last row of Tab.~\ref{tab:counting_models}, excluding the vector new physics, we find 75 models up to 3 BSM particles and 14 models with 2 BSM particles.
It is notable that all models without vectors contain at most 3 BSM particles. In other words, this means that all models with 4 BSM particles, which generate the operator classes $\Psi^4 H^2 D$ and $\Psi^2 H^4 (D/W)^2$, necessarily contain vectors. In the following, we focus on the 14 minimal models without vectors, which can yield interesting patterns, and leave non-minimal UV completions with 3 or 4 BSM particles for future work. 

\subsection{Minimal models with dimension-9 contribution to $0\nu\beta\beta$ and loop-suppressed Weinberg operator}
\label{sec:minimal_models}

\begin{table}
\centering
\footnotesize
\begin{tabular}{llllll}
\toprule
\# & Model & dim-9 tree level & dim-7 1L & dim-5 & type \\
\midrule
M1 & ($S_{8,2,\frac{1}{2}}, F_{8,1,0}$) & $\mathcal{O}^{(9)}_{dQdQLL}$, $\mathcal{O}^{(9)}_{dQQuLL}$, & $\mathcal{O}^{(7)}_{LH}$, $\mathcal{O}^{(7)}_{LHD}$, $\mathcal{O}^{(7)}_{LLduD}$, & 1L & T3 \\
& & $\mathcal{O}^{(9)}_{QuQuLL}$ & $\mathcal{O}^{(7)}_{LLQdH}$, $\mathcal{O}^{(7)}_{LLQuH}$ & & \\
\cmidrule{1-6}
M2 & ($S_{8,2,\frac{1}{2}}, F_{8,3,0}$) & $\mathcal{O}^{(9)}_{dQdQLL}$, $\mathcal{O}^{(9)}_{dQQuLL}$, & $\mathcal{O}^{(7)}_{LH}$, $\mathcal{O}^{(7)}_{LHD}$, $\mathcal{O}^{(7)}_{LLduD}$,  & 1L & T3 \\
& & $\mathcal{O}^{(9)}_{QuQuLL}$ & $\mathcal{O}^{(7)}_{LLQdH}$, $\mathcal{O}^{(7)}_{LLQuH}$ & & \\
\cmidrule{1-6}
M3 & ($S_{1,1,1}, S_{1,3,0}$) & $\mathcal{O}^{(9)}_{LLH^4 (D/W)^2}$ & $\mathcal{O}^{(7)}_{LHD}$ & 2L, anti & CLBZ-10 \\
\cmidrule{1-6}
M4 & ($S_{1,1,1}, S_{8,2,\frac{1}{2}}$) & $\mathcal{O}^{(9)}_{dQdQLL}$, $\mathcal{O}^{(9)}_{dQQuLL}$, & $\mathcal{O}^{(7)}_{LLduD}$, $\mathcal{O}^{(7)}_{LLQdH}$, & 2L, anti & NG \\
& & $\mathcal{O}^{(9)}_{QuQuLL}$ & $\mathcal{O}^{(7)}_{LLQuH}$ & & \\
\cmidrule{1-6}
M5 & ($S_{3,1,-\frac{1}{3}}, S_{3,1,\frac{2}{3}}$) &  $\mathcal{O}^{(9)}_{ddueue}$, $\mathcal{O}^{(9)}_{dQdueL}$, & $\mathcal{O}^{(7)}_{LLQdH}$, $\mathcal{O}^{(7)}_{LeudH}$ & 2L, anti & CLBZ-1 \\
& & $\mathcal{O}^{(9)}_{dQdQLL}$ & & & \\
\cmidrule{1-6}
M6 & ($S_{3,1,-\frac{1}{3}}, S_{6,1,-\frac{2}{3}}$) &  $\mathcal{O}^{(9)}_{ddueue}$, $\mathcal{O}^{(9)}_{dQdueL}$, & $\mathcal{O}^{(7)}_{LLQdH}$, $\mathcal{O}^{(7)}_{LeudH}$ & 2L, sym & CLBZ-1 \\
& & $\mathcal{O}^{(9)}_{dQdQLL}$ & & & \\
\cmidrule{1-6}
M7 & ($S_{3,3,-\frac{1}{3}}, S_{3,1,\frac{2}{3}}$) & $\mathcal{O}^{(9)}_{dQdQLL}$ & $\mathcal{O}^{(7)}_{LLQdH}$, $\mathcal{O}^{(7)}_{LeudH}$ & 2L, anti & CLBZ-1 \\
\cmidrule{1-6}
M8 & ($S_{3,3,-\frac{1}{3}}, S_{6,1,-\frac{2}{3}}$) & $\mathcal{O}^{(9)}_{dQdQLL}$ & $\mathcal{O}^{(7)}_{LLQdH}$, $\mathcal{O}^{(7)}_{LeudH}$ & 2L, sym & CLBZ-1 \\
\cmidrule{1-6}
M9 &($S_{3,2,\frac{1}{6}}, S_{6,1,\frac{1}{3}}$) &  $\mathcal{O}^{(9)}_{dQdQLL}$ & $\mathcal{O}^{(7)}_{LLduD}$, $\mathcal{O}^{(7)}_{LLQdH}$, & 2L, anti & CLBZ-1 \\
& &  & $\mathcal{O}^{(7)}_{LeudH}$ & & \\
\cmidrule{1-6}
M10 & ($S_{3,2,\frac{1}{6}}, S_{6,3,\frac{1}{3}}$) &  $\mathcal{O}^{(9)}_{dQdQLL}$ & $\mathcal{O}^{(7)}_{LLQdH}$, $\mathcal{O}^{(7)}_{LeudH}$ & 2L, anti & CLBZ-1 \\
\cmidrule{1-6}
M11 & ($S_{3,1,-\frac{1}{3}}, F_{8,1,0}$) &  $\mathcal{O}^{(9)}_{ddueue}$, $\mathcal{O}^{(9)}_{dQdueL}$, & $\mathcal{O}^{(7)}_{LLQdH}$, $\mathcal{O}^{(7)}_{LeudH}$ & 2L, sym & PTBM-1 \\
& & $\mathcal{O}^{(9)}_{dQdQLL}$ & & & \\
\cmidrule{1-6}
M12 & ($S_{3,3,-\frac{1}{3}}, F_{8,3,0}$) & $\mathcal{O}^{(9)}_{dQdQLL}$ & $\mathcal{O}^{(7)}_{LLQdH}$ & 2L, sym & PTBM-1 \\
\cmidrule{1-6}
M13 & ($S_{3,2,\frac{1}{6}}, F_{8,1,0}$) & $\mathcal{O}^{(9)}_{dQdQLL}$ & $\mathcal{O}^{(7)}_{LLQdH}$ & 2L, sym & PTBM-1 \\
\cmidrule{1-6}
M14 & ($S_{3,2,\frac{1}{6}}, F_{8,3,0}$) & $\mathcal{O}^{(9)}_{dQdQLL}$ & $\mathcal{O}^{(7)}_{LLQdH}$ & 2L, sym & PTBM-1 \\
\bottomrule
\end{tabular}
\caption{Classification of minimal models and the LNV dimension-9 operators they induce at tree-level and the dimension-7 operators they give at 1-loop level. The penultimate column indicates whether the model leads to the Weinberg operator at 1-loop or 2-loop order, and whether the 2-loop contributions are symmetric (sym) or antisymmetric (anti); see the text for details. The last column classifies the diagrams for the Weinberg operator according to~\cite{Bonnet_2012} for 1-loop topologies and based on~\cite{AristizabalSierra:2014wal} for 2-loop topologies. }
\label{tab:classification_minimal}
\end{table}

\begin{figure}[h!]
    \centering
\includegraphics[scale=0.9]{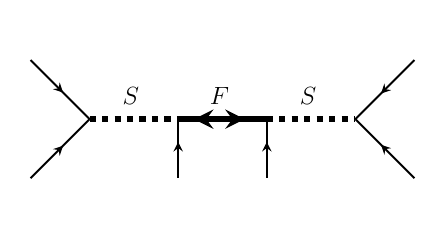}
\includegraphics[scale=0.9]{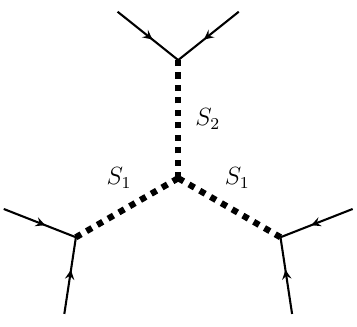}
    \caption{Diagrams for the dimension-9 operators for the minimal models. All models with one BSM scalar $S$ and one BSM vectorlike fermion $F$ generate operators of the class $\Psi^6$ via the diagram on the left. Models with two BSM scalars $S_1$ and $S_2$ generate $\Psi^6$ operators via the diagram on the right.}    
\label{fig:diagrams_dim9}
\end{figure}

In Tab.~\ref{tab:classification_minimal}, we list all 14 minimal models with their particle content in the second column. As we focus on scalar and fermionic SM extensions, the models can consist of either a single scalar and a single fermion or two scalars. The third column lists all dimension-9 operator structures produced by each model at the tree level. Except for M3, all other models generate operators of the class $\Psi^6$, i.e., operators with 6 SM fermions, as listed in Tab.~\ref{tab:operators_dim9}. These operators are opened by the diagrams shown in Fig.~\ref{fig:diagrams_dim9}, on the left for one scalar and one fermion and on the right for two scalars. Note that here the term \textit{diagrams} refers to the second step in the diagram-based approach presented above, and that in order to obtain specific model diagrams, one has to insert all possible combinations of the external fermions into these diagrams and get the viable quantum numbers for the internal BSM particles. For example, if we choose two external legs that meet at one vertex to be ingoing $l$ and $q$, then the ingoing BSM scalar has to be a colour $\bar{3}$, an $SU(2)$ singlet or triplet, and of hypercharge $1/3$.

\begin{figure}[t!]
    \centering
    \includegraphics[scale=0.9]{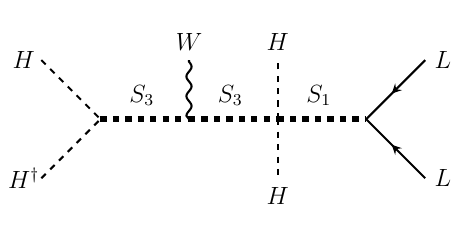}
    \hskip5mm
    \includegraphics[scale=0.9]{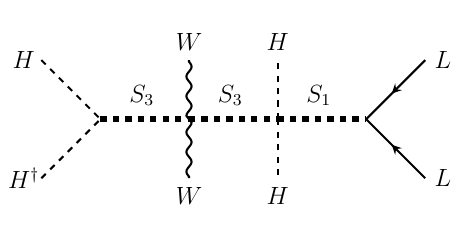}
    \caption{Tree-level diagrams for dimension-9 operator $\mathcal{O}^{(9)}_{LLH^4D^2}$ with one $W$ boson (left) and two $W$ bosons (right) for the model M3. Note that the operator with no $W$ boson, i.e.~with only partial derivatives $\partial_{\mu}$, vanishes due to symmetries. }
    \label{fig:diagrams_M3}
\end{figure}

The model M3 matches onto the operator structure $\mathcal{O}_{LLH^4 (D/W)^2}$ instead of 6-fermion operators, and the model diagrams are given separately in Fig.~\ref{fig:diagrams_M3}. The field strength tensor as well as the covariant derivative contain partial derivatives and gauge fields, such that the operator structure can contain the terms $\partial_{\mu} \partial^{\mu}$, $\partial_{\mu} W^{\mu}$ and $W_{\mu} W^{\mu}$. In Fig.~\ref{fig:diagrams_M3} we show the model diagrams with one $W$ and two $W$s. The diagram with $\partial_{\mu} \partial^{\mu}$ vanishes because of $SU(2)$ antisymmetry -- we effectively contract an $SU(2)$ singlet $S_1$ with three identical doublets $H$ and another doublet $H^{\dagger}$. This contraction is antisymmetric in the three Higgs fields, and hence vanishes for identical fields.

In the same manner, the fourth column in Tab.~\ref{tab:classification_minimal} lists the dimension-7 operators that are generated at 1-loop level. As stated in Tab.~\ref{tab:operator_structures}, there are 5 different classes, and the realisation of the diagrams depends on the type of operator. We will provide the diagrams explicitly for the example models that we discuss later in detail. 

\begin{figure}[h!]
    \centering
\includegraphics[scale=1]{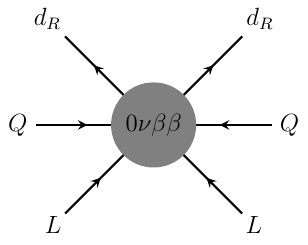}
\includegraphics[scale=1]{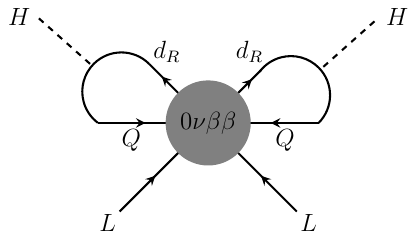}
    \caption{Black box theorem for the Weinberg operator: every model that generates operator $\mathcal{O}_{dQdQLL}^{(9)}$ at tree level (left) will generate the Weinberg operator at maximally 2-loop level by closing both pairs of external $Q$ and $d_R$ legs into a loop via a Yukawa interaction with a Higgs boson (right).}    
\label{fig:blackbox_theorem}
\end{figure}

Next, the penultimate column states whether the Weinberg operator is produced at 1-loop (1L), or 2-loop (2L) level.
Note that since all models except for M3 generate dimension-9 operator $\mathcal{O}^{(9)}_{dQdQLL}$ at tree-level, the black box theorem states that the Weinberg operator is produced at maximally 2-loop order. This can be easily seen by closing each pair of external $Q$ and $d_R$ lines into loops with Higgs insertion, cf.~Fig.~\ref{fig:blackbox_theorem}.

We furthermore give a symmetry flag ``sym" for a symmetric and ``anti" for an antisymmetric flavour structure in the Weinberg operator,~i.e.~$C_{LH}^{(5)pr} =  C_{LH}^{(5)rp}$ or $C_{LH}^{(5)pr} = - C_{LH}^{(5)rp}$. Clearly, this translates to a symmetric and antisymmetric neutrino mass matrix, respectively. Note that only symmetric mass matrices can give rise to physical neutrino masses. Since $0\nu\beta\beta$ involves only first-generation quarks and leptons, this also implies that models with antisymmetric contributions to the Weinberg operator will not trigger a dimension-5 contribution to $0\nu\beta\beta$. We will focus on models with symmetric neutrino mass matrices in the following.

To determine the dominant contribution to $0\nu\beta\beta$ for all classes of models, it will be crucial to understand whether and at which order the models contribute to neutrino masses. The last column in~Tab.~\ref{tab:classification_minimal} contains the topology type of the Weinberg diagram, which is useful to group the models in classes. For the models that produce the Weinberg operator at 1 loop, i.e.~M1 and M2, we follow the convention of~\cite{Bonnet_2012} and identify them as type T3. The diagrams for this class of models will be given explicitly in Sec.~\ref{sec:1L_model}. For the models that generate the Weinberg operator at 2 loops, we follow~\cite{AristizabalSierra:2014wal}. 
Models M5-M10 contain two scalars and lead to a double-box topology of type CLBZ-1 from~\cite{AristizabalSierra:2014wal}. As will be explained in detail in Sec.~\ref{sec:2L_model}, models M5, M7 and M9 give antisymmetric contributions to the Weinberg operator due to antisymmetric colour or $SU(2)$ contractions. Models M6, M8, and M10 lead to viable symmetric neutrino mass matrices at the 2-loop level.

Models M11-M14 contain one BSM scalar and one BSM fermion and give a double-box topology of type PTBM-1~\cite{AristizabalSierra:2014wal}. In this class, all four models lead to symmetric neutrino mass matrices.
All 2-loop Weinberg diagrams are listed in Appendix~\ref{sec:2loopWein}. Specifically, the CLBZ-1-type models are shown in 
Fig.~\ref{fig:CLBZ1} and the PTBM-1-type models are depicted in Fig.~\ref{fig:PTBM1}.

\begin{figure}[h!]
    \centering
\includegraphics[scale=1]{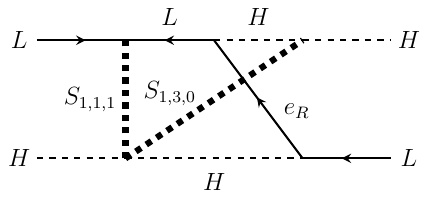}
\includegraphics[scale=1]{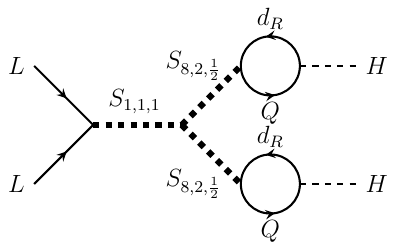}
    \caption{Two-loop diagrams for the Weinberg operator in models M3 (left) and M4 (right). Note that both models give only antisymmetric contributions to the Weinberg operator.}    
\label{fig:Weinberg_M3+M4}
\end{figure}

The two remaining models, M3 and M4, are rather peculiar, because they contain the scalar $S_{1,1,1}$, which couples to two lepton doublets $L$. The $SU(2)$ contraction of 2 doublets with a singlet is antisymmetric in the lepton flavours. This implies that, already at tree level, only one of the leptons can be electron-flavoured, and hence neither model will contribute to $0\nu\beta\beta$ at tree level. We list them regardless, because they might be interesting for other LNV experimental probes, such as $\mu$ to $e$ conversion, which will be left for future work. M3 generates the Weinberg operator via the CLBZ-10 2-loop topology in~\cite{AristizabalSierra:2007nf}, cf.~the diagram on the left in Fig.~\ref{fig:Weinberg_M3+M4}. Since the vertex $L^p L^r S_{1,1,1}$ is antisymmetric in lepton flavours $p$ and $r$ and the two internal vertices with SM electron Yukawa couplings can be assumed flavour diagonal, the whole diagram is necessarily antisymmetric in the external lepton flavours. For M4 one can draw 2-loop diagrams for the Weinberg operator, cf.~e.g.~the right diagram in Fig.~\ref{fig:Weinberg_M3+M4}. These diagrams, however, are not genuine 2-loop diagrams given that the loops lead to effective couplings between the SM Higgs and the BSM scalar $S_{8,2,\tfrac{1}{2}}$; hence, we classify them as non-genuine (NG). Both external leptons couple directly to $S_{1,1,1}$ as in the tree-level case and the antisymmetry is evident.

In Tab. \ref{tab:operator_to_model} we show explicitly which dimension-9, dimension-7 and dimension-5 operators are generated by which of the minimal models. This grid representation makes it easy to identify all models that can contribute to a specific operator. 

\renewcommand{\arraystretch}{1.5}
\begin{table}[h]
\centering
\footnotesize
\addtolength{\tabcolsep}{-0.3em}
\begin{tabular}{lcccccccccccccc}
\toprule
\#     & M1 & M2 & M3 & M4 & M5 & M6 & M7 & M8 & M9 & M10 & M11 & M12 & M13 & M14 \\
Model &
\rot{\left( S_{8,2,\frac{1}{2}}, F_{8,1,0}\right)} &
\rot{\left(S_{8,2,\frac{1}{2}}, F_{8,3,0}\right)} &
\rot{\left(S_{1,1,1}, S_{1,3,0}\right)} &
\rot{\left(S_{1,1,1}, S_{8,2,\frac{1}{2}}\right)} &
\rot{\left(S_{3,1,-\frac{1}{3}}, S_{3,1,\frac{2}{3}}\right)} &
\rot{\left(S_{3,1,-\frac{1}{3}}, S_{6,1,-\frac{2}{3}}\right)} &
\rot{\left(S_{3,3,-\frac{1}{3}}, S_{3,1,\frac{2}{3}}\right)} &
\rot{\left(S_{3,3,-\frac{1}{3}}, S_{6,1,-\frac{2}{3}}\right)} &
\rot{\left(S_{3,2,\frac{1}{6}}, S_{6,1,\frac{1}{3}}\right)} &
\rot{\left(S_{3,2,\frac{1}{6}}, S_{6,3,\frac{1}{3}}\right)} &
\rot{\left(S_{3,1,-\frac{1}{3}}, F_{8,1,0}\right)} &
\rot{\left(S_{3,3,-\frac{1}{3}}, F_{8,3,0}\right)} &
\rot{\left(S_{3,2,\frac{1}{6}}, F_{8,1,0}\right)} &
\rot{\left(S_{3,2,\frac{1}{6}}, F_{8,3,0}\right)} \\
\midrule
$\mathcal{O}^{(9)}_{ddueue}$        &  &  &  &  & T & T &  &  &  &  & T &  &  &  \\
$\mathcal{O}^{(9)}_{dQdueL}$        &  &  &  &  & T & T &  &  &  &  & T &  &  &  \\
$\mathcal{O}^{(9)}_{dQdQLL}$        & T & T &  & T & T & T & T & T & T & T & T & T & T & T \\
$\mathcal{O}^{(9)}_{dQQuLL}$        & T & T &  & T &  &  &  &  &  &  &  &  &  &  \\
$\mathcal{O}^{(9)}_{QuQuLL}$        & T & T &  & T &  &  &  &  &  &  &  &  &  &  \\
$\mathcal{O}^{(9)}_{LLH^4(D/W)^2}$  &  &  & T &  &  &  &  &  &  &  &  &  &  &  \\
\hline
$\mathcal{O}^{(7)}_{LH}$            & 1L & 1L &  &  &  &  &  &  &  &  &  &  &  &  \\
$\mathcal{O}^{(7)}_{LHD}$           & 1L & 1L & 1L &  &  &  &  &  &  &  &  &  &  &  \\
$\mathcal{O}^{(7)}_{LLduD}$         & 1L & 1L &  & 1L &  &  &  &  & 1L &  &  &  &  &  \\
$\mathcal{O}^{(7)}_{LLQdH}$         & 1L & 1L &  & 1L & 1L & 1L & 1L & 1L & 1L & 1L & 1L & 1L & 1L & 1L \\
$\mathcal{O}^{(7)}_{LLQuH}$         & 1L & 1L &  & 1L &  &  &  &  &  &  &  &  &  &  \\
$\mathcal{O}^{(7)}_{LeudH}$         &  &  &  &  & 1L & 1L & 1L & 1L & 1L & 1L & 1L &  &  &  \\
\hline
$\mathcal{O}^{(5)}_{LH}$            & 1L & 1L & \makecell{2L\\anti} & \makecell{2L\\anti} & \makecell{2L\\anti} & \makecell{2L\\sym} & \makecell{2L\\anti} & \makecell{2L\\sym} & \makecell{2L\\anti} & \makecell{2L\\sym} & \makecell{2L\\sym} & \makecell{2L\\sym} & \makecell{2L\\sym} & \makecell{2L\\sym} \\
%type                                 & \rottext{T3} & \rottext{T3} & \rottext{CLBZ-10} & \rottext{NG} & \rottext{CLBZ-1} & \rottext{CLBZ-1} & \rottext{CLBZ-1} & \rottext{CLBZ-1} & \rottext{CLBZ-1} & \rottext{CLBZ-1} & \rottext{PTBM-1} & \rottext{PTBM-1} & \rottext{PTBM-1} & \rottext{PTBM-1} \\
\bottomrule
\end{tabular}
\caption{Operator-to-model table for the 14 minimal models in Tab.~\ref{tab:classification_minimal}. The entries T, 1L and 2L denote that the model in the given column generates the operator in the corresponding row at tree, 1-loop, or 2-loop level, respectively. The flag 'sym'/'anti' refers to symmetric/antisymmetric neutrino mass matrices, see text for details.}
\label{tab:operator_to_model}
\end{table}

In Sec.~\ref{sec:1L_model} we will discuss in detail the models which generate the Weinberg operator at 1 loop (M1 and M2), calculating the matching of the full UV theory to the SMEFT and studying the contributions of the dimension-5, dimension-7 and dimension-9 operators to $0\nu\beta\beta$. In Sec.~\ref{sec:2L_model} we will present a thorough study of the models that generate the Weinberg operator at 2 loops and we will focus on models that can trigger $0\nu\beta\beta$ at tree-level and lead to symmetric neutrino mass matrices, i.e.~M5, M7, M9 and M11-M14.

\section{Minimal models with 1-loop generated Weinberg operator}
\label{sec:1L_model}
The two models in this class are M1 with
$S_{8,2,\frac{1}{2}}$ and $F_{8,1,0}$ and M2 with $S_{8,2,\frac{1}{2}}$ and $F_{8,3,0}$. Note that the only difference between these models is the $SU(2)$ multiplicity of the heavy vector-like fermion, which will result in a difference of order 1 in the matching due to group factors. Since we aim to compare the orders of magnitude of $0\nu\beta\beta$ half-lives, we proceed with the first model, M1, and note that all qualitative statements should also hold for M2.

We use the shorthand notation $S_8\equiv S_{8,2,\frac{1}{2}}$ for the scalar colour octet and $F_8 \equiv F_{8,1,0}$ for the fermionic colour octet. 
The interaction Lagrangian is given by

\begin{equation}
\begin{split}
\mathcal{L}_{Int}^{UV} = & y_{SQd}^{pr} T_b^{Aa} S_8^{Ai} \left(\overline{Q}_{ai}^p  P_R  d^{br} \right) + y_{SQu}^{pr} T_b^{Aa} \epsilon^{ij} \bar{S}_{8,j}^A \left(\overline{Q}_{ai}^p  P_R  u^{br} \right) \\
& + y_{SFL}^p \bar{\epsilon}_{ij} S_8^{Aj} \left( F_8^{A,T}  C P_L  L^{ip} \right) + \lambda_{S} H^i H^j \bar{S}_{8,i}^A \bar{S}_{8,j}^A + \text{h.c.}, \\
\end{split}
\end{equation}
where $T^A$ denote the 8 generators of $SU(3)$ and $\epsilon$ the 2-dimensional completely antisymmetric tensor with the convention $\epsilon^{12} = 1$. We denote by $y_{SQd}$ and $y_{SQu}$ the Yukawa-like couplings of $S_8$ to a quark doublet $Q$ and a right-handed down-type ($d$) or up-type ($u$) quark singlet, respectively. The coupling of $S_8$ to the vector-like fermion $F_8$ and the lepton doublet $L$ is given by $y_{SFL}$, and $\lambda_S$ refers to the quartic scalar coupling of two instances of $S_8$ and two Higgs doublets. \\

\subsection{Matching to the SMEFT}
As it can be seen in Tab.~\ref{tab:classification_minimal}, this model generates the dimension-9 operators $\mathcal{O}_{dQdQLL}^{(9)}$, $\mathcal{O}_{dQQuLL}^{(9)}$ and $\mathcal{O}_{QuQuLL}^{(9)}$ at tree level, and the dimension-7 operators $\mathcal{O}_{LHD}^{(7)}$, $\mathcal{O}_{LLduD}^{(7)}$, $\mathcal{O}_{LLQdH}^{(7)}$ and $\mathcal{O}_{LLQuH}^{(7)}$ as well as the dimension-5 Weinberg operator at 1-loop level. 

We use \texttt{Matchete} \cite{Fuentes-Martin:2022jrf} to calculate the matching of the model to the dimension-9, dimension-7 and dimension-5 operators up to 1-loop level. Since \texttt{Matchete} provides the matching results in an arbitrary Green's basis, we then transform the relevant results to our basis by applying integration-by-parts (IBP), equations of motion (EOM) and Fierz identities. Furthermore, for simplicity, we assume a common mass scale for the new physics, $m_{S_8} = m_{F_8} = \Lambda $~\footnote{For a hierarchical new physics scenario, a two-step matching will be necessary and there will be a logarithmic correction involving the ratio of the two heavy scales, see e.g.~\cite{Fridell:2024pmw} for some relevant discussion.}.

\begin{figure}[t!]
    \centering
    \begin{subfigure}[t]{\textwidth}
    \includegraphics[scale=0.63]{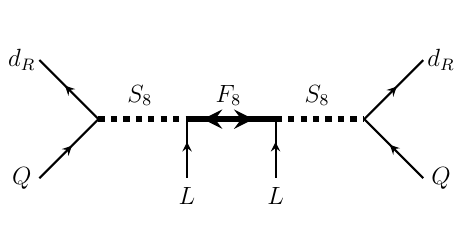}
    \includegraphics[scale=0.63]{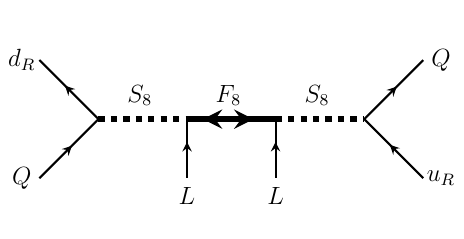}
    \includegraphics[scale=0.63]{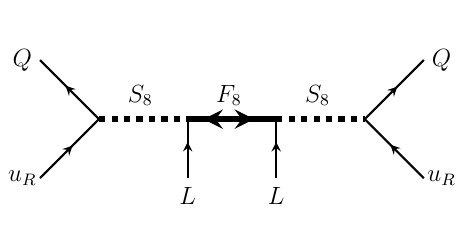}
    \caption{Tree-level diagrams for the dimension-9 operators $\mathcal{O}_{dQdQLL}^{(9)}$, $\mathcal{O}_{dQQuLL}^{(9)}$ and $\mathcal{O}_{QuQuLL}^{(9)}$}
    \end{subfigure}
    \begin{subfigure}[t]{\textwidth}
    \centering
     \includegraphics[scale=0.63]{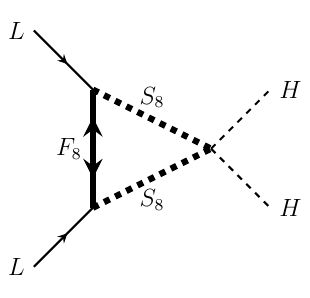}
    \includegraphics[scale=0.63]{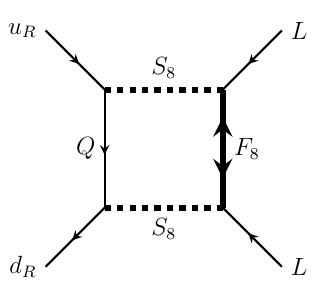} \\
    \includegraphics[scale=0.63]{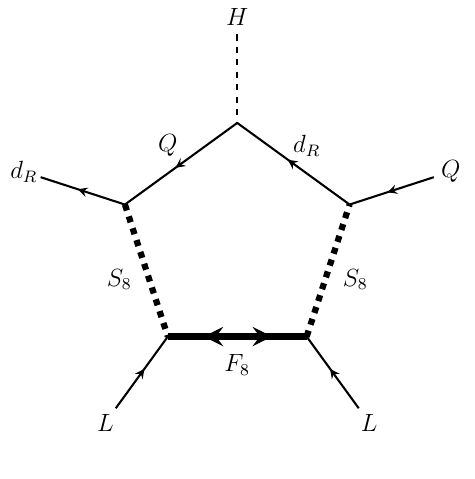}
    \includegraphics[scale=0.63]{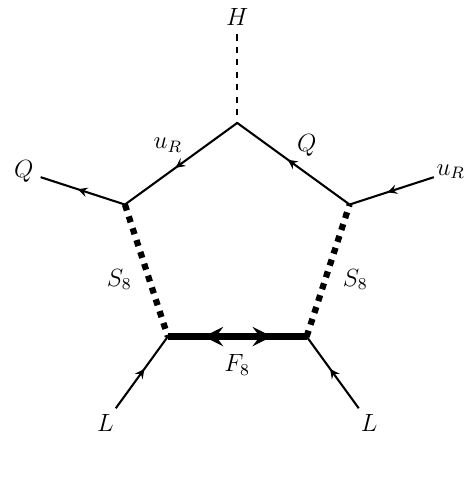}
    \caption{1-loop diagrams for the dimension-7 operators $\mathcal{O}_{LHD}^{(7)}$, $\mathcal{O}_{LLduD}^{(7)}$, $\mathcal{O}_{LLQdH}^{(7)}$ and $\mathcal{O}_{LLQuH}^{(7)}$}
    \end{subfigure}
    \begin{subfigure}[t]{\textwidth}
    \centering
    \includegraphics[scale=0.63]{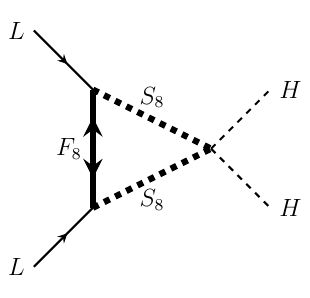}
    \caption{1-loop diagram for the Weinberg operator}
    \end{subfigure}
    \caption{Dimension-9, dimension-7 and dimension-5 diagrams generated by the model M1 containing $S_{8,2,\frac{1}{2}}$ and $F_{8,1,0}$.}
    \label{fig:M1_diagrams}
\end{figure}

Due to the structure of this model, all tree-level interactions at dimension 9 violate lepton number by 2 units. For the matching of the diagrams in Fig.\ \ref{fig:M1_diagrams} (a) expressed in the dimension-9 basis in \cite{Liao:2020jmn}, we find
\begin{equation}
\begin{split}
    c_{dQdQLL1}^{(9),\, prstuv} & = - \frac{1}{4 \sqrt{2}} \frac{1}{\Lambda^5}y_{SFL}^u y_{SFL}^v \left(y_{SQd}^{rp}\right)^* \left(y_{SQd}^{ts}\right)^* \,, \\ 
    c_{dQdQLL2}^{(9), \, prstuv} & =  \frac{3}{4 \sqrt{2}} \frac{1}{\Lambda^5}y_{SFL}^u y_{SFL}^v \left(y_{SQd}^{rp}\right)^* \left(y_{SQd}^{ts}\right)^* \,, \\
     c_{dQQuLL1}^{(9), \, prstuv} & =  \frac{1}{2 \sqrt{2}} \frac{1}{\Lambda^5} y_{SFL}^u y_{SFL}^v \left(y_{SQd}^{rp}\right)^* y_{SQu}^{st} \,, \\ 
    c_{dQQuLL2}^{(9), \, prstuv} & =  -\frac{3}{2 \sqrt{2}} \frac{1}{\Lambda^5} y_{SFL}^u y_{SFL}^v \left(y_{SQd}^{rp}\right)^* y_{SQu}^{st} \,, \\
    c_{QuQuLL1}^{(9), \, prstuv} & = - \frac{1}{4 \sqrt{2}} \frac{1}{\Lambda^5}y_{SFL}^u y_{SFL}^v y_{SQu}^{pr} y_{SQu}^{st} \,, \\ 
    c_{QuQuLL2}^{(9), \, prstuv} & =  \frac{3}{4 \sqrt{2}} \frac{1}{\Lambda^5}y_{SFL}^u y_{SFL}^v y_{SQu}^{pr} y_{SQu}^{st}\,. \\
\end{split}
\end{equation}

The model also generates dimension-6 operators at tree level, but by construction $S_8$ couples to a pair of quark and anti-quark, and thus all generated dimension-6 interactions conserve baryon number. The same applies to the dimension-8 operators at tree-level.

Next, we obtain 1-loop contributions to the  dimension-7 operators $\mathcal{O}_{LHD}^{(7)}$, $\mathcal{O}_{LLduD}^{(7)}$, $\mathcal{O}_{LLQdH}^{(7)}$ and $\mathcal{O}_{LLQuH}^{(7)}$ as shown in Fig.\ \ref{fig:M1_diagrams} (b). Transforming the operators into the dimension-7 basis~\cite{Lehman:2014jma} and reading off the Wilson coefficients yields

\begin{equation}
\begin{split}
    c_{LHD1}^{(7), \, pr} &= \frac{2}{3} \frac{1}{16 \pi^2}\frac{1}{\Lambda^3} y_{SFL}^p y_{SFL}^r \lambda_S \,,\\
    c_{LHD2}^{(7),\, pr} &= -\frac{4}{3} \frac{1}{16 \pi^2}\frac{1}{\Lambda^3} y_{SFL}^p y_{SFL}^r \lambda_S \,, \\
    c_{LLduD1}^{(7),\, prst} &= \frac{i}{3\sqrt{2}} \frac{1}{16 \pi^2}\frac{1}{\Lambda^3} y_{SFL}^p y_{SFL}^r \left(y_{SQd}^{us}\right)^* y_{SQu}^{ut} \,, \\
    c_{LLQdH1}^{(7), \, prst} &= 0 \,, \\
    c_{LLQdH2}^{(7), \, prst} &= -\frac{4}{\sqrt{2}} \frac{1}{16 \pi^2}\frac{1}{\Lambda^3} y_{SFL}^p y_{SFL}^r \left(y_{SQd}^{su}\right)^* \left(y_{SQd}^{vt}\right)^* y_d^{vu} \,, \\
    c_{LLQuH}^{(7), \, prst} &= \frac{2}{\sqrt{2}} \frac{1}{16 \pi^2}\frac{1}{\Lambda^3} y_{SFL}^p y_{SFL}^r y_{SQu}^{sv} y_{SQu}^{ut} \left(y_u^{uv}\right)^* \,.
\end{split}
\end{equation}

Finally, this model generates the Weinberg operator at the 1-loop level, as it contains a quartic scalar coupling of $S_8$ to $H$, via the diagram in Fig.\ \ref{fig:M1_diagrams} (c) and the matching reads
\begin{equation}
    c_{LH}^{(5),\, pr} = -\frac{1}{4\pi^2} \frac{1}{\Lambda} \lambda_S y_{SFL}^p y_{SFL}^r \,.
    \label{eq:M2_1L_Weinberg}
\end{equation}

\subsection{Neutrino masses and neutrinoless double beta decay}

Since the Yukawa-like couplings $y_{SFL}$ enter the dimension-9 and dimension-5 matching in exactly the same way, their size will affect the rate estimations of $0\nu\beta \beta$, but not qualitatively the ratio of the two contributions. Hence, we do not aim at fitting the neutrino masses precisely. Instead, we constrain the couplings entering the Weinberg operator in Eq.\ (\ref{eq:M2_1L_Weinberg}) by imposing that $M_{\nu\nu} \lesssim 2.5 \cdot 10^{-3}$~eV. With $M_{\nu\nu} = -v^2 c_{LH}^{(5)}$, we find for $\Lambda = 2.5$~TeV, the lowest possible mass scale for the leptoquark not excluded by direct collider searches~\cite{Schmaltz:2018nls}, the upper limit $(y_{SFL})^2 \lambda_S \sim 10^{-12}$.  The other Yukawa-type couplings $Y_{SQd}$ and $Y_{SQu}$ are unconstrained by the neutrino masses and enter the dimension-9 and dimension-7 operators in the same form. Therefore, we can assume they are of order one. We furthermore assume that only the first generation couplings are nonzero, $y_{SQd}^{11} = y_{SQu}^{11} = 1$, i.e.\ all quarks propagating in the loops for the dimension-7 operators are of first generation.

We analyse two benchmark scenarios 
\begin{equation}
\begin{split}
    \emph{B1}\text{: } &y_{SFL} = 10^{-4}, \quad \lambda_S = 10^{-4} \,, \\
    \emph{B2}\text{: } &y_{SFL} = 1,  \qquad \, \, \,\lambda_S = 10^{-12} \,.
\end{split}
\end{equation}
In the first scenario, the couplings of interest are equal in size, which could be motivated by both emerging from new physics at the same mass scale. The second scenario suppresses $\lambda_S$ with respect to the natural $y_{SFL}$ and is motivated by a setup in which the quartic scalar coupling $\lambda_S$ is symmetry-protected.

\begin{figure}
    \centering
    \begin{subfigure}[t]{0.49\textwidth}
         \includegraphics[width=\linewidth]{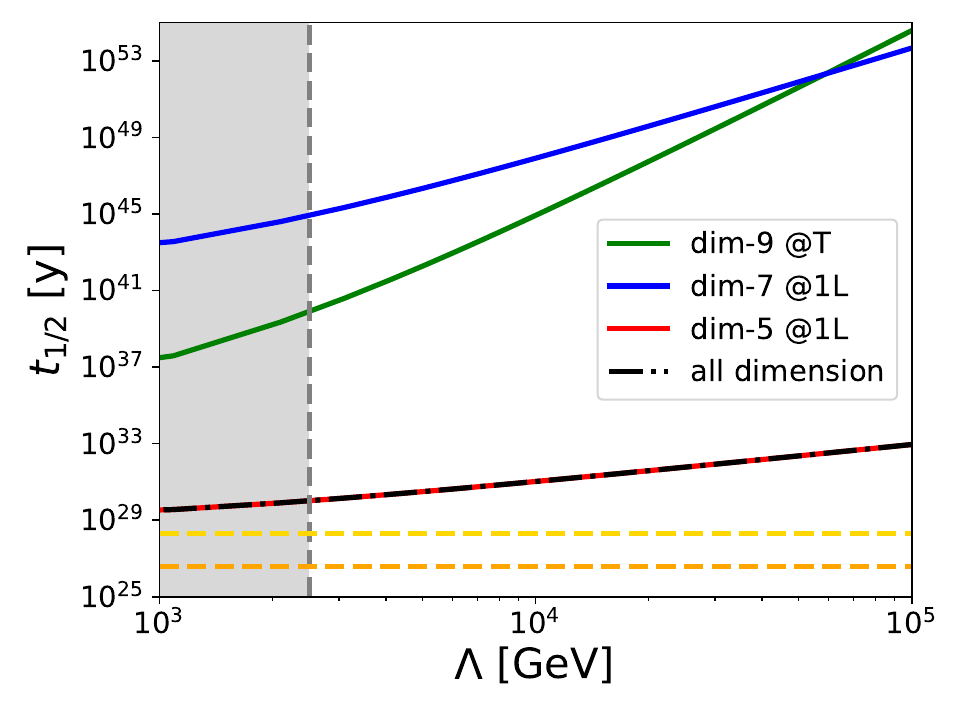}
         \caption{\emph{B1}: $y_{SFL}^1 = 10^{-4}$, $\lambda_S = 10^{-4}$}
    \end{subfigure}
      \begin{subfigure}[t]{0.49\textwidth}
         \includegraphics[width=\linewidth]{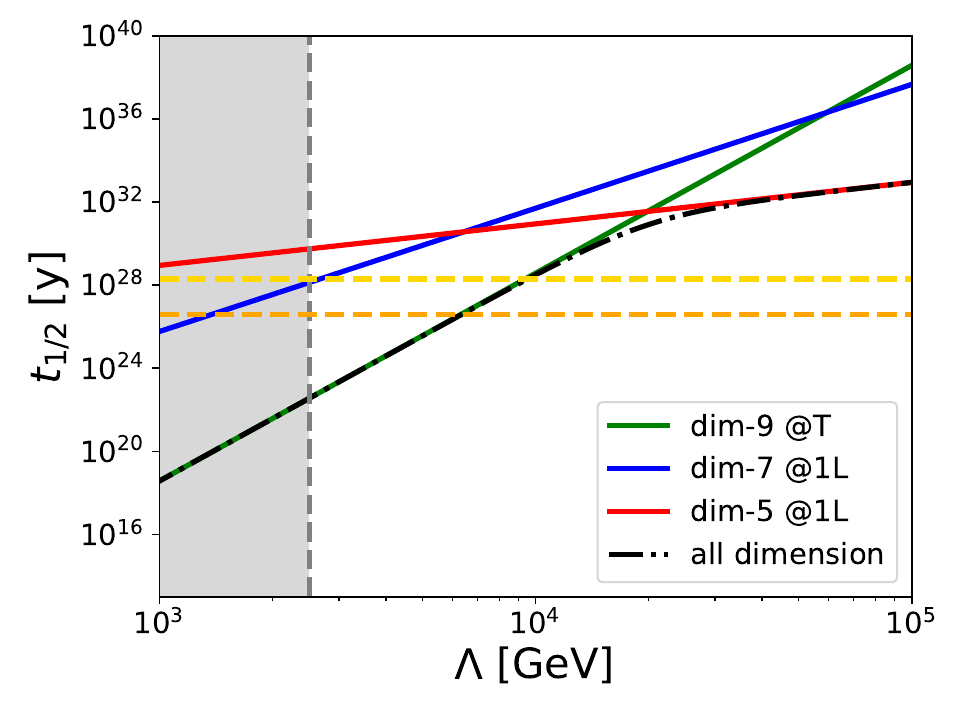}
         \caption{\emph{B2}: $y_{SFL}^1 = 1$, $\lambda_S = 10^{-12}$ }
    \end{subfigure}
    \caption{The $0 \nu\beta \beta$ half-life in $^{136}$Xe induced by dimension-9 (green), dimension-7 (blue) and dimension-5 (red) operators, respectively. The black dashed-dotted line shows the combined contribution of all operators at a time, the orange dotted line the current exclusion limit from KamLAND-Zen ($3.8 \cdot 10^{26}$ years) and the yellow dashed lines the expected future sensitivity ($2 \cdot 10^{28}$ years). Masses below $2.5$ TeV are excluded from direct searches for leptoquarks and shaded in gray. The subplots show two scenarios that can explain the correct order of neutrino masses: on the left, with couplings of similar size, and on the right, with a small quartic scalar coupling; see the text for details. }
    \label{fig:M2_half_lives}
\end{figure}

To calculate the contribution of the SMEFT operators to $0\nu\beta\beta$ we make use of the python tool \texttt{$\nu$DoBe} \cite{Scholer:2023bnn}. It takes the SMEFT Wilson coefficients and the corresponding scale as inputs and computes the $0\nu\beta\beta$ rate, including matching and (Low-energy Effective Field Theory) RGE running. We use the half-life of the isotope $^{136}$Xe to compare the impact of the different operators, employing the nuclear matrix elements calculated in IBM-2 nuclear structure model~\cite{Graf:2018ozy,Deppisch:2020ztt}. Since we are interested in an order-of-magnitude estimate, we adopt the default assumption and set the unknown low-energy constants (LECs) to zero. In Appendix~\ref{sec:LECs} we examine the importance of the unknown LECs on $0\nu\beta\beta$ for the example model in Sec.~\ref{sec:2L_model}.

We use the relation $m_{\beta\beta} = -\frac{1}{2} C_{LH}^{(5) ee}v^2$. In principle, $m_{\beta\beta}$ would also receive contributions from the dimension-7 operator $\mathcal{O}_{LH}^{(7)}$, but it is not produced in this model. 
Furthermore, we focus on pure matching effects and neglect SMEFT operator mixing through RGE running~\cite{Graf:2025cfk}.
For example, other dimension-7 operators could mix at 1-loop level into $\mathcal{O}^{(7)}_{LH}$. But since the dimension-7 operators themselves are produced at 1-loop level, the effective contribution would be a 2-loop effect, which is expected to be subdominant with respect to the direct matching contribution to the Weinberg operator. 

We compute the half-life contributions in $^{136}$Xe for all three classes of operators, dimension-9, dimension-7 and dimension-5, individually and for all operators together as a function of the new physics scale $\Lambda$. The results are shown in Fig.~\ref{fig:M2_half_lives} for the two benchmarks \emph{B1} (left) and \emph{B2} (right). The grey shaded area $\Lambda < 2.5$ TeV is excluded by direct searches for leptoquarks \cite{Schmaltz:2018nls} while the orange and yellow dashed line show the current KamLAND-Zen limit of $3.8 \cdot 10^{26}$ years \cite{KamLAND-Zen:2024eml} and the future projected nEXO sensitivity of $2 \cdot 10^{28}$ years \cite{nEXO:2021ujk}, respectively. In the first benchmark, the dimension-5 contributions will dominate over the dimension-7 and dimension-9 contributions at all mass scales.
In the second benchmark, the dimension-9 contributions are sufficiently enhanced and dominate over the dimension-5 and dimension-7 operators for $\Lambda \lesssim 20$ TeV. For $6 \text{ TeV} \lesssim \Lambda \lesssim 10 \text{ TeV}$, the half-life predicted by the dimension-9 operators lies in between the current exclusion limit and the future sensitivity and can be probed in the next-generation $0\nu\beta\beta$ experiments.

%%%%%%%%%%%%%%%%%%%%%%%%%%%%%%%%%%%%%%%%%%
%%%%%%%%%%%%%%%%%%%%%%%%%%%%%%%%%%%%%%%%%%
\section{Minimal models with 2-loop generated Weinberg operator} 
\label{sec:2L_model}
%%%%%%%%%%%%%%%%%%%%%%%%%%%%%%%%%%%%%%%%%%
\subsection{General implications for the Weinberg operator}
\label{sec:2L_classification}

\begin{figure}
    \begin{subfigure}[t]{0.49\textwidth}
         \includegraphics[width=\linewidth]{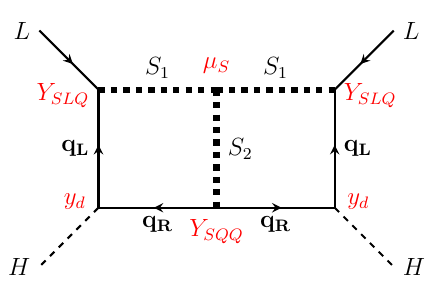}
         \caption{generic CLBZ1}
    \end{subfigure}
    \begin{subfigure}[t]{0.49\textwidth}
         \includegraphics[width=\linewidth]{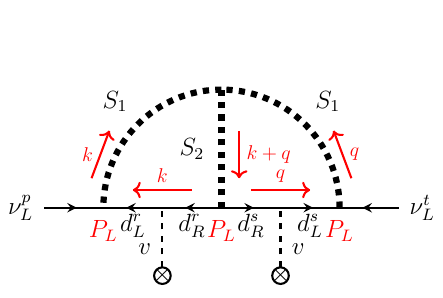}
         \caption{neutrino masses from CLBZ1}
    \end{subfigure}
    \begin{subfigure}[t]{0.49\textwidth}
         \includegraphics[width=\linewidth]{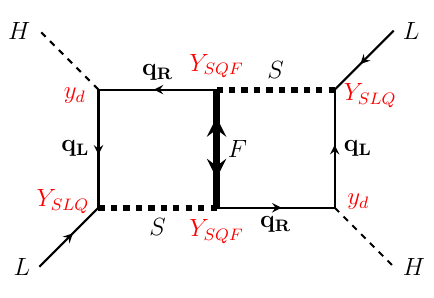}
         \caption{generic PTBM1}
    \end{subfigure}
    \begin{subfigure}[t]{0.49\textwidth}
         \includegraphics[width=\linewidth]{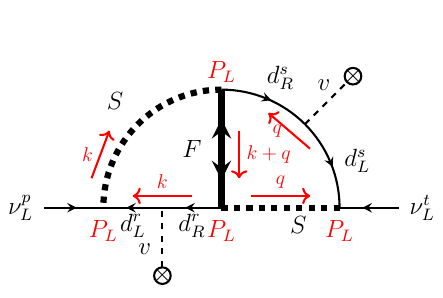}
         \caption{neutrino masses from PTBM1}
    \end{subfigure}
    \caption{Generic CLBZ-1 and PTBM-1 2-loop diagrams for the Weinberg operator and the corresponding neutrino masses. }
    \label{fig:generic_CLBZ1_PTBM1}
\end{figure}

As shown in Tab.~\ref{tab:classification_minimal}, the models M5-M10 give rise to CLBZ-1 2-loop Weinberg topologies and the models M11-M14 to PTBM-1 topologies. We first show that all these models lead to a similar expression for the neutrino mass matrix.

We denote by $\mathbf{q_L}$ a left-handed generic quark which could be either $Q$ or $\bar{d}_R$ and by 
$\mathbf{q_R}$ a right-handed quark $d_R$ or $\bar{Q}$. 
The corresponding couplings of a leptoquark to $L$ and $\mathbf{q_L}$ are denoted by $Y_{SLQ}$. In the PTBM-1 case, the coupling of $S$ to $\mathbf{q_R}$ and the vector-like fermion $F$ is $Y_{SQF}$ and in the CLBZ-1 case the coupling of $S_2$ to a pair of $\mathbf{q_R}$ and $\mathbf{q_R}$ is $Y_{SQQ}$. Furthermore, $y_d$ stands for the SM down-type Yukawa coupling and $\mu_S$ for a triple scalar coupling in the CLBZ-1 case.

Since we evaluate the diagram in the broken phase, the insertion of the vacuum expectation value flips the chirality in the down-type quark propagator.
We choose the convention that $y_d$ is diagonal, i.e.\ all propagating down-type quarks coincide with the mass eigenstates, and we do not have to include any CKM rotations.

In the PTBM-1 case, the neutrino mass matrix gives
\begin{equation}
    M_{\nu\nu}^{pt} \propto \sum_{r,s=1}^{3} Y_{SLQ}^{pr} Y_{SLQ}^{ts} Y_{SQF}^r Y_{SQF}^s I_{PTBM}^{rs}(m_S, m_F)\,,
\end{equation}
where the proportionality denotes that, for individual models, the precise prefactor will differ due to group factors coming from the color and $SU(2)$ contractions at each vertex, which are of order 1 to 10. The loop factor is given by 
\begin{equation}
    I_{PTBM}^{rs} =  \int \frac{d^4 k}{(2\pi)^4} \int \frac{d^4 q}{(2\pi)^4} \frac{m_{d_r} m_{d_s} m_F}{(k^2-m_{d_r}^2) (k^2-m_S^2) (q^2-m_{d_s}^2) (q^2 - m_S^2) ((k+q)^2- m_F^2)} \,.
\end{equation}
The mass terms in the numerator arise from the chirality projectors projecting out the masses in the fermion propagators. Without loss of generality, we can set $Y_{SQF}^{r}=1$ since we can always reabsorb it into a redefinition of $Y_{SLQ}^{pr}$.

In the CLBZ-1 case, we find the neutrino mass matrix (up to proportionality) to be
\begin{equation}
    M_{\nu\nu}^{pt} \propto \sum_{r,s=1}^{3} Y_{SLQ}^{pr} Y_{SLQ}^{ts} Y_{SQQ}^{rs} I_{CLBZ}^{rs}(m_S, m_F),
   \label{eq:Mnunu_CLBZ}
\end{equation}
with 
\begin{equation}
    I_{CLBZ}^{rs} =  \int \frac{d^4 k}{(2\pi)^4} \int \frac{d^4 q}{(2\pi)^4} \frac{ \mu_S m_{d_r} m_{d_s}}{(k^2-m_{d_r}^2) (k^2-m_S^2) (q^2-m_{d_s}^2) (q^2 - m_S^2) ((k+q)^2- m_F^2)}. 
\end{equation}
Note that, since the innermost propagator is a scalar rather than a fermion, we do not obtain a mass insertion in the numerator. On the other hand, the triple scalar coupling $\mu_S$ is dimensionful and, in meaningful UV models, associated with the new-physics scale, potentially through a vacuum expectation value generating the heavy new-physics masses. Identifying $m_F \sim \mu_S \sim \Lambda$, the loop factors for both CLBZ-1 and PTBM-1 are exactly the same. 

The Weinberg diagrams for M5-M10 are shown in Fig.\ \ref{fig:CLBZ1} and for M11-M14 in Fig.\ \ref{fig:PTBM1} in the Appendix. For the CLBZ-1 models we can distinguish two cases: the diagrams for M5, M7 and M9 contain an antisymmetric vertex $S_2 \mathbf{q_R}\mathbf{q_R}$ due to the antisymmetric colour contractions in M5 and M7 and the antisymmetric $SU(2)$ contractions in M9. This means that the coupling $Y_{SQQ}^{rs}$ is antisymmetric in the quark flavours $r$ and $s$. Since the loop factor $I_{CLBZ}^{rs}$ is fully symmetric, it is easy to see from Eq.~\ref{eq:Mnunu_CLBZ} that the diagonal contributions to $M_{\nu\nu}$ vanish and that these models do not give neutrino masses.

For the other three models, M6, M8, and M10, the couplings $Y_{SQQ}$ are symmetric, yielding a symmetric neutrino mass matrix. In these cases, we can set $Y_{SQQ}^{rs} =1$, or similarly absorb it into $Y_{SLQ}$ as it should factor into a vector product $Y_{SQ}^r \times Y_{SQ}^s$.

This means that all four PTBM-1 diagrams and all three symmetric CLBZ-1 diagrams give exactly the same estimate for the neutrino masses up to group factors of order one to ten. In the following subsections, we will explicitly work out the matching, the fit to the neutrino masses and mixing angles, and the expected $0\nu\beta\beta$ rate for an example model. Based on the results, we conclude that the qualitative findings for this model apply to all the models discussed here.

\subsection{Matching to the SMEFT for $S_{3,1,-\frac{1}{3}}$, $F_{8,1,0}$}
We choose to work out the model M11, which contains a scalar leptoquark $S_3 \equiv S_{3,1,-\frac{1}{3}}$ and a fermionic colour octet $F_8 \equiv F_{8,1,0}$ and the interaction Lagrangian reads 
\begin{equation}
\begin{split}
\mathcal{L}_{\text{int}}^{\text{UV}} = & \, y_{Sue}^{pr} S_3^a \left( \overline{u_a^p} \cdot e^{r,C} \right) + y_{SLQ}^{pr}  \epsilon^{ij}S_3^a \left( \overline{Q_{ai}^p} \cdot L_j^{r,C} \right) + y_{SdF}^p T_b^{Aa} S_3^b \left(\overline{d_a^p}\cdot P_L \cdot F_8^A\right) + \text{h.c.},
\end{split}
\end{equation}
where $T^A$ denote the 8 generators of $SU(3)$ and we adopt the normalisation $T^A = \frac{\sqrt{3}}{2 \sqrt[4]{2}} \lambda^A$ with the Gell-Mann matrices $\lambda^A$. Note that a different convention can be reabsorbed in a redefinition of $Y_{SdF}$. The leptoquark couples to the singlets $e_R$ and $u_R$ via the Yukawa-like couplings $y_{Sue}$, and to the doublets $L$ and $Q$ via $y_{SLQ}$. As it will become clear later, the couplings to the doublets, $y_{SLQ}$, give the dominant contribution to $0\nu\beta\beta$ and, in contrast to $Y_{Sue}$, can be constrained by neutrino masses. $y_{SdF}$ denotes the Yukawa-like coupling of $S_3$ to $d_R$ and $F_8$.

The full Lagrangian could, in principle, also contain a diquark coupling of the leptoquark
\begin{equation}
\mathcal{L}_{\text{int}}^{\text{UV}} \supset y_{Sud}^{pr} \overline{S_{3,c}} f^{abc} \left(\overline{u_a^p} \cdot d_b^{r,C} \right) + \text{h.c.}
\end{equation}
This term would induce baryon-number-violating dimension-6 operators, which would predict proton decays. The limits on the proton lifetime imply that $y_{Sud}$ has to be tiny, and we can neglect the diquark coupling in the following.

\begin{figure}[t!]
    \centering
    \begin{subfigure}[t]{\textwidth}
    \includegraphics[scale=0.63]{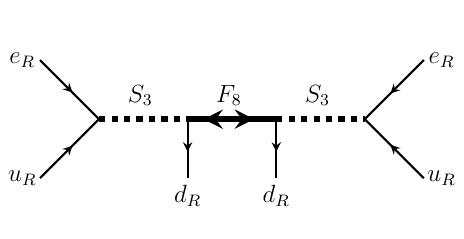}
    %\hskip5mm
    \includegraphics[scale=0.63]{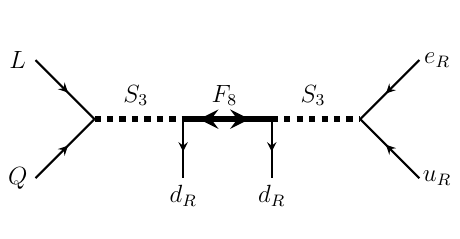}
    \includegraphics[scale=0.63]{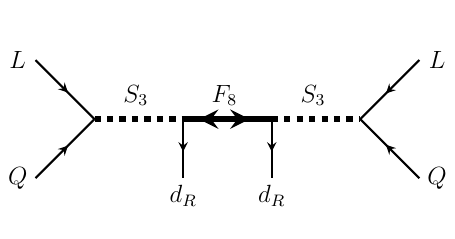}
    \caption{Tree-level diagrams for the dimension-9 operators $\mathcal{O}^{(9)}_{ddueue}$, $\mathcal{O}^{(9)}_{dQdueL}$ and $\mathcal{O}^{(9)}_{dQdQLL}$}
    \end{subfigure}
    \begin{subfigure}[t]{\textwidth}
    \centering
    \includegraphics[scale=0.63]{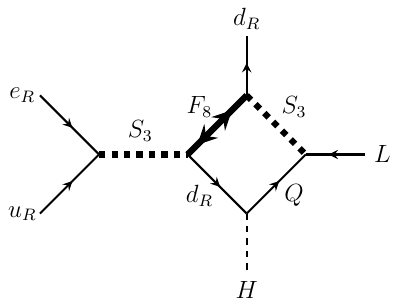}
    %\hskip5mm
    \includegraphics[scale=0.63]{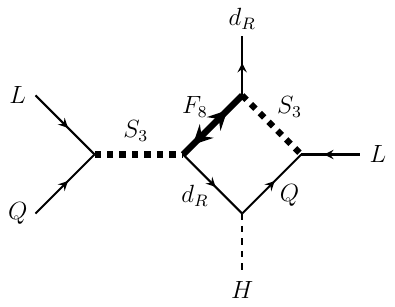}
    \caption{1-loop diagrams for the dimension-7 operators $\mathcal{O}^{(7)}_{LeudH}$ and $\mathcal{O}^{(7)}_{LLQdH}$}
    \end{subfigure}
    \begin{subfigure}[t]{\textwidth}
    \centering
    \includegraphics[scale=0.63]{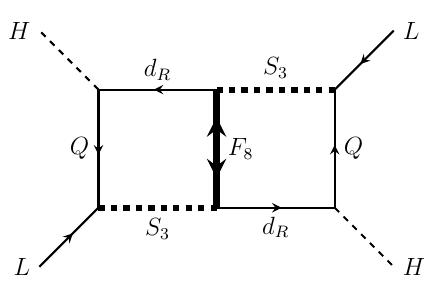}
    \caption{2-loop double-box diagram for the Weinberg operator}
    \end{subfigure}
    \caption{Dimension-9, dimension-7 and dimension-5 diagrams for model M11 with $S_{3,1,-\frac{1}{3}}$ and $F_{8,1,0}$.}
    \label{fig:M11_diagrams}
\end{figure}

First, we calculate the tree-level matching to the dimension-9 operators and the 1-loop matching for the dimension-7 operators using \texttt{Matchete}. As can be seen in Fig.~\ref{tab:classification_minimal}, this model produces the dimension-9 operators $\mathcal{O}_{ddueue}^{(9)}$, $\mathcal{O}_{dQdueL1}^{(9)}$, $\mathcal{O}_{dQdueL2}^{(9)}$, $\mathcal{O}_{dQdQLL1}^{(9)}$ and $\mathcal{O}_{dQdQLL2}^{(9)}$ through the diagrams in Fig.~\ref{fig:M11_diagrams} (a). We use various IBP and Fierz identities and equations of motion to transform the operator structures obtained in \texttt{Matchete} into our dimension-9 basis. We give only the matching for the operators in Tab.~\ref{tab:operators_dim9} that contribute at tree-level to $0\nu\beta\beta$. Assuming the same mass for both $S_3$ and $F_8$, $m_{S_3} = m_{F_8} = \Lambda$, we find 
\begin{equation}
\begin{split}
    c_{ddueue}^{(9), \, prstuv} &= \frac{1}{2\sqrt{2}} \frac{1}{\Lambda^5} y_{SdF}^p y_{SdF}^r \left(y_{Sue}^{st} \right)^* \left(y_{Sue}^{uv}\right)^*, \\
    c_{dQdueL1}^{(9), \, prstuv} &= \frac{1}{2\sqrt{2}} \frac{1}{\Lambda^5} y_{SdF}^p y_{SdF}^s \left( y_{SLQ}^{vr}\right)^* \left( y_{Sue}^{tu}\right)^*, \\
    c_{dQdueL2}^{(9), \, prstuv} &= \frac{1}{2\sqrt{2}} \frac{1}{\Lambda^5} y_{SdF}^p y_{SdF}^s \left( y_{SLQ}^{vr}\right)^* \left(y_{Sue}^{tu}\right)^*, \\
    c_{dQdQLL1}^{(9), \, prstuv} &= \frac{1}{4\sqrt{2}} \frac{1}{\Lambda^5} y_{SdF}^p y_{SdF}^s \left(y_{SLQ}^{ur}\right)^* \left( y_{SLQ}^{vt}\right)^*, \\
    c_{dQdQLL2}^{(9), \, prstuv} &= \frac{1}{4\sqrt{2}} \frac{1}{\Lambda^5} y_{SdF}^p y_{SdF}^s \left(y_{SLQ}^{ur}\right)^* \left(y_{SLQ}^{vt}\right)^*.
\end{split}
\end{equation}

The dimension-7 operators $\mathcal{O}_{LeudH}^{(7)}$, $\mathcal{O}_{LLQdH,1}^{(7)}$ and $\mathcal{O}_{LLQdH,2}^{(7)}$ are produced at 1-loop, cf.\ the diagrams in Fig.\ \ref{fig:M11_diagrams} (b). As implied by the external particles, the dimension-7 1-loop contributions are obtained from the dimension-9 diagrams by closing a $Q$ and a $d_R$ to a loop via a Higgs insertion. Again, we use various identities to transform the \texttt{Matchete} output into our dimension-7 basis and give the matching results for the operators in Tab.~\ref{tab:operators_dim7}, which contribute to $0\nu \beta \beta$ at tree-level. The matching is given by
\begin{equation}
\begin{split}
    c_{LeudH}^{(7),\, prst} &= -\frac{1}{16 \pi^2} \frac{ \sqrt{2}}{\Lambda^3}  y_{SdF}^t y_{SdF}^u \left(y_{Sue}^{sr}\right)^* \left(y_{SLQ}^{pv}\right)^* y_d^{vu}, \\
    c_{LLQdH1}^{(7),\, prst} & = -\frac{1}{16 \pi^2} \frac{ 2\sqrt{2}}{\Lambda^3} y_{SdF}^t y_{SdF}^u \left(y_{SLQ}^{rs}\right)^* \left(y_{SLQ}^{pv}\right)^* y_d^{vu}, \\
    c_{LLQdH2}^{(7),\, prstuv} & = \frac{1}{16 \pi^2} \frac{ 2\sqrt{2}}{\Lambda^3} y_{SdF}^t y_{SdF}^u \left(y_{SLQ}^{rs}\right)^* \left(y_{SLQ}^{pv}\right)^* y_d^{vu}.
\end{split}
\end{equation}
Closing the loop with two SM quarks introduces the down-type Yukawa matrix.

Finally, as discussed in general terms in Sec.~\ref{sec:2L_classification}, this model produces the Weinberg operator at 2-loop level via the diagram shown in Fig.~\ref{fig:M11_diagrams} (c), which can be seen by closing the remaining $Q$ and $d_R$ legs of the 1-loop diagrams for the dimension-7 operator $\mathcal{O}^{(7)}_{LLQdH}$. Here, we work out the precise contribution of this model to the neutrino masses. Since \texttt{Matchete} currently works up to 1-loop, we match the Weinberg operator at 2-loop by hand, making use of the master integrals in~\cite{AristizabalSierra:2014wal}. 

\begin{figure}[t!]
    \centering
    \includegraphics[scale=1]{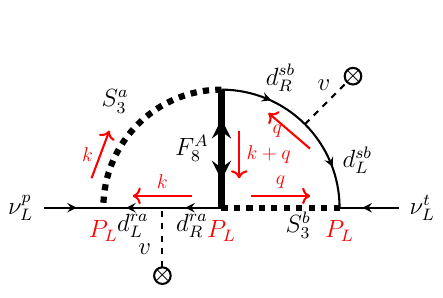}
    \caption{2-loop diagram for the neutrino masses in the broken phase with momentum assignment for model M11 with $S_{3,1,-\frac{1}{3}}$ and $F_{8,1,0}$.}
\label{fig:M11_diagram_neutrino_masses}
\end{figure}

Calculating the matrix element following the diagram in Fig.\ \ref{fig:M11_diagram_neutrino_masses} gives 
\begin{equation}
    i \mathcal{M}^{pt} = - 6 \sqrt{2} i m_{d^r} m_{d^s} m_{F_8} \left( y_{SLQ}^{pr} \right)^* \left( y_{SLQ}^{ts} \right)^* y_{SdF}^r y_{SdF}^s \overline{\nu_{L}^{C,p}} \nu_L^t \mathcal{I}_{d^r d^s, S_3 S_3, F_8} \,,
\end{equation}
with the colour sum $\sum_{A} Tr \left( T^A T^A \right) = 6 \sqrt{2}$ and the scalar integral
\begin{equation}
    \mathcal{I}_{ab,\alpha \beta, X} = \int \frac{d^4k}{\left(2\pi\right)^4} \int \frac{d^4q}{\left(2\pi\right)^4} \frac{1}{\left(k^2 - m_a^2\right) \left(q^2 - m_b^2\right) \left(k^2 - m_{\alpha}^2\right) \left(q^2 - m_{\beta}^2\right) \left((k+q)^2 - m_X\right)^2}.
\end{equation}
Its solution can be expressed as
\begin{equation}
    \mathcal{I}_{ab,\alpha \beta,X} = \frac{1}{(4\pi)^4} \frac{1}{m_X^2} \frac{1}{(t_{\alpha} -r_a) (t_{\beta} - r_b)}\Big[-\hat{g}(t_{\alpha}, t_{\beta})+ \hat{g}(t_{\alpha}, r_b) + \hat{g}(r_a, t_{\beta}) -\hat{g}(r_a, r_b)
    \Big] ,
    \label{eq:I_definition}
\end{equation}
and the mass ratios are given by
\begin{equation}
    r_a = \left(\frac{m_{d^r}}{m_{F_8}}\right)^2, \qquad r_b = \left(\frac{m_{d^s}}{m_{F_8}}\right)^2, \qquad t_{\alpha} = \left(\frac{m_{S_3}}{m_{F_8}}\right)^2, \qquad t_{\beta} = \left(\frac{m_{S_3}}{m_{F_8}}\right)^2.
\end{equation}
The finite part of the solution to the master integral yields
\begin{equation}
\begin{split}
  \hat{g}(s,t) = & \frac{s}{2} \ln(s) \ln(t) + \sum_{\pm} \pm \frac{s(1-s) + 3st + 2(1-t)x_{\pm}}{2 \omega} \\ & \times  \Big[ \text{Li}_2\left( \frac{x_{\pm}}{x_{\pm}-s} \right)- \text{Li}_2\left( \frac{x_{\pm}-s}{x_{\pm}}\right) +
  \text{Li}_2\left( \frac{t-1}{x_{\pm}} \right) - \text{Li}_2\left( \frac{t-1}{x_{\pm}-s} \right) \Big],
\end{split}
\end{equation}
where 
\begin{equation}
    x_{\pm} = \frac{1}{2} \left( -1 + s +t \pm \omega \right), \qquad \omega = \sqrt{1 + s^2 + t^2 - 2(s+t+st)},
\end{equation}
and the dilogarithm is given by
\begin{equation}
    \text{Li}_{2} = - \int_0^x \frac{\ln(1-y)}{y} dy.
\end{equation}
The matrix element translates into the matching for the Weinberg operator,
\begin{equation}
    c_{LH}^{(5)pt} =  -6 \sqrt{2} y_d^r y_d^s m_{F_8} \left( y_{SLQ}^{pr} \right)^* \left( y_{SLQ}^{ts} \right)^* y_{SdF}^r y_{SdF}^s \mathcal{I}_{d^r d^s, S_3 S_3, F_8}.
\end{equation} 
In the following, we numerically evaluate the model and try to find points in the parameter space that
\begin{enumerate}
    \item[(i)] correctly reproduces the neutrino masses and mixing;
    \item[(ii)] are compatible with the experimental limits from experiments probing charged-lepton-flavour violation;
    \item[(iii)] yield a contribution to the $0\nu\beta\beta$ rate induced by the dimension-9 operators that dominates over the loop-suppressed dimension-5 and dimension-7 contributions;
    \item[(iv)] give a $0\nu\beta\beta$ half-life above the current exclusion limits but below the future sensitivity, such that the model could be probed in the upcoming searches.
\end{enumerate}

\subsection{Fit to neutrino masses for one generation of $S_3$}
We start by fitting the neutrino masses and assume a minimal particle content in the model, with only one generation of both $S_3$ and $F_8$.
We extract the neutrino mass matrix $M_{\nu\nu}$ from the matrix element
\begin{equation}
    \mathcal{M}^{pt} = -M_{\nu\nu}^{pt} \overline{\nu_{L}^{C,p}} \nu_L^t,
\end{equation}
and assume that it can be diagonalised to
\begin{equation}
    D_{\nu} = \begin{pmatrix}
        m_{\nu}^1 & 0 & 0 \\ 0 & m_{\nu}^2 & 0 \\ 0 & 0 & m_{\nu}^3  
    \end{pmatrix},
\end{equation}
via
\begin{equation}
    D_{\nu} = U^T M_{\nu\nu} U .
\label{eq:Dnu}
\end{equation}
Here, the unitary matrix,
\begin{equation}
    U = U_{PMNS} \begin{pmatrix} 1 & 0 & 0 \\ 0 & e^{-i\alpha/2} & 0  \\ 0 & 0 & e^{-i\beta/2}
    \end{pmatrix},
\end{equation}
incorporates the PMNS matrix
\begin{equation}
    U_{PMNS} = \begin{pmatrix}
        1 & 0 & 0 \\ 0 & c_{23} & s_{23} \\ 0 & -s_{23} & c_{23}
    \end{pmatrix} 
    \begin{pmatrix}
        c_{13} & 0 & s_{13} e^{-i \delta} \\ 0 & 1 & 0 \\ -s_{13} e^{i \delta} & 0 & c_{13} 
    \end{pmatrix}
    \begin{pmatrix}
        c_{12} & s_{12} & 0 \\ -s_{12} & c_{12} & 0 \\ 0 & 0 & 1
    \end{pmatrix},
\end{equation}
with $c_{ij} \equiv \cos \left( \Theta_{ij} \right)$, $s_{ij} \equiv \sin \left( \Theta_{ij} \right)$,
and two additional complex phases $\alpha$ and $\beta$. We adopt a phase convention where the neutrino masses are taken to be positive real numbers, and any leftover phases from diagonalising the mass matrix are moved into two “Majorana phases” ($\alpha$, $\beta$), while the remaining CP-violating phase is kept as the usual Dirac phase $\delta$ inside the PMNS matrix.
Assuming normal mass ordering and the lightest eigenstate to be massless, the masses read
\begin{equation}
    \begin{split}
        m_{\nu}^1 &= 0, \\
        m_{\nu}^2 &= \sqrt{\Delta m_{12}^2}, \\
        m_{\nu}^3 &= \sqrt{\Delta m_{13}^2}.\\
    \end{split}
\end{equation}
The mass splittings $\Delta m_{12}^2$ and $\Delta m_{13}^2$ as well as the 3 mixing angles $\Theta_{12}$, $\Theta_{13}$ and $\Theta_{23}$ and the complex phase $\delta$ are determined by fitting the experimental data~\cite{deSalas:2020pgw,Esteban:2020cvm,ParticleDataGroup:2024cfk}. 

In order to probe the values for $y_{SLQ}$ that are compatible with the observed neutrino masses and mixing, we follow the Casas-Ibarra parametrisation~\cite{Casas:2001sr}~\footnote{See also~\cite{Herrero-Garcia:2025aox} for a generalisation to nonsymmetric radiative neutrino masses.}. We define the loop function $I$ as
\begin{equation}
    I^{rs} = 6 \sqrt{2} m_{F_8} m_d^r m_d^s y_{SdF}^r y_{SdF}^s \mathcal{I}_{d^r d^s, S_3 S_3, F_8},
\end{equation}
such that
\begin{equation}
    M_{\nu\nu}^{pt} = \left(y_{SLQ}^{ps}\right)^* \left(y_{SLQ}^{tr}\right)^*  I^{rs}, 
\label{eq:Mnunu_LF}
\end{equation}
or in matrix form
\begin{equation}
    M_{\nu\nu} = y_{SLQ}^* \, I \, y_{SLQ}^{\dagger}. 
\end{equation}

In the following, we will set $y_{SdF}^{r} =1$ to evaluate the loop factor. Note that this is equivalent to a redefinition of $\left(\tilde{y}_{SLQ}^{ps}\right)^* = \left(y_{SLQ}^{ps}\right)^* y_{SdF}^s$ (no summation over s), i.e.,~we effectively obtain bounds on the product of the Yukawa couplings whenever we constrain $y_{SLQ}$ in the following.

Note that the loop factor $I$ needs to be of rank-3 for the above equation to render three independent neutrino masses. This means that we can not restrict ourselves to only third-generation quarks propagating in the loop. If $I$ is a complex symmetric rank-3 matrix, we can always diagonalise it via
\begin{equation}
    I_D = S^T I S,
\end{equation}
where $I_D$ is a real, non-negative diagonal matrix and $S$ stands for a unitary matrix with $S^{\dagger} S =\mathbb{1}$, i.e.\ $S^{-1} = S^{\dagger}$.
Solving for $I$ gives the expression for the neutrino mass matrix
\begin{equation}
    M_{\nu\nu} = y_{SLQ}^* S^* I_D  S^{\dagger} y_{SLQ}^{\dagger}.
\label{eq:Mnunu_LFdiag}
\end{equation}
Inserting Eq.\ (\ref{eq:Mnunu_LFdiag}) into Eq.\ (\ref{eq:Dnu}) then yields
\begin{equation}
    D_{\nu}=  U^T y_{SLQ}^* S^* I_D S^{\dagger} y_{SLQ}^{\dagger} U.
\end{equation}
Multiplying this expression by $\sqrt{D_{\nu}^{-1}}$ from both sides gives
\begin{equation}
    \mathbb{1} = \Big[ \sqrt{I_D} S^{\dagger} \, y_{SLQ}^{\dagger} U \sqrt{D_{\nu}^{-1}} \Big]^T \Big[ \sqrt{I_D} \, S^{\dagger} y_{SLQ}^{\dagger} U \sqrt{D_{\nu}^{-1}} \Big].
    \label{eq:CI_unity}
\end{equation}
This implies that
\begin{equation}
    \mathcal{R} \equiv \sqrt{I_D} S^{\dagger} y_{SLQ}^{\dagger} U \sqrt{D_{\nu}^{-1}} 
\label{eq:R_definition}
\end{equation}
is an orthogonal $3\times 3$ matrix with $\mathcal{R}^T \mathcal{R} = \mathbb{1}$ which can be parametrised in its general form by
\begin{equation}
    \mathcal{R} = \begin{pmatrix}
        1 & 0 & 0 \\ 0 & \cos \Theta_a & \sin \Theta_a \\ 0 & -\sin \Theta_a & \cos \Theta_a
    \end{pmatrix} 
    \begin{pmatrix}
        \cos \Theta_b & \sin \Theta_b & 0 \\ -\sin \Theta_b & \cos \Theta_b & 0 \\ 0 & 0 & 1
    \end{pmatrix}
    \begin{pmatrix}
        \cos \Theta_c & 0 & \sin \Theta_c  \\ 0 & 1 & 0 \\ -\sin \Theta_c  & 0 & \cos \Theta_c 
    \end{pmatrix}\,.
    \label{eq:R_parametrisation}
\end{equation}

Given that a rank-3 loop factor renders $I_D$ invertible, we can solve Eq.\ (\ref{eq:R_definition}) for the Yukawa couplings
\begin{equation}
    y_{SLQ} = U \sqrt{D_{\nu}} \mathcal{R}^T \sqrt{\left(I_D^{-1}\right)^*} S^{\dagger}
    \label{eq:YSLQ_CI}.
\end{equation}
The diagonal loop factor $I_D$ can, in principle, be complex, but we found the imaginary part to be negligible in all cases.
Keeping all 3 quark generations in the loop factor is necessary to give all 9 entries of the Yukawa matrix $y_{SLQ}$. Now we can sample $y_{SLQ}$ from Eq.\ (\ref{eq:YSLQ_CI}): while $D_{\nu}$ and $U_{PMNS}$ are fully determined by the experiment, the free parameters are the scales $m_S$ and $m_F$ in $I$, the two complex phases $\alpha$ and $\beta$ in $U$, and the three real mixing angles $\Theta_a$, $\Theta_b$ and $\Theta_c$ the orthogonal matrix $\mathcal{R}$ can be parametrised by. For fixed $m_S$ and $m_F$ we then generate three thousand random samples for $\alpha$, $\beta$, $\Theta_a$, $\Theta_b$ and $\Theta_c$ (all within the interval $[0, 2 \pi]$) and we keep the samples for which the absolute value of all entries in $y_{SLQ}$ is perturbative, i.e.~$y_{SLQ}^{pr} < \sqrt{4 \pi}$.

Taking into account the full loop factor with all quark generations, we find that for $m_S = m_F = \Lambda = 5$ TeV the eigenvalues of the loop factor read
\begin{equation}
    I_D = \begin{pmatrix}
        -3.4 \cdot 10^{-19} & 0 & 0 \\
        0 & 2.3 \cdot 10^{-11} & 0 \\
        0 & 0 & 2.2 \cdot 10^3 
    \end{pmatrix} \mathrm{eV} \,.
\end{equation}
To compensate for the huge hierarchy in $I_D$, Eq.\ (\ref{eq:YSLQ_CI}) dictates an inverse hierarchy in the Yukawa couplings among the quark generations, $y_{SLQ}^{p1} \gg y_{SLQ}^{p2} \gg y_{SLQ}^{p3}$.

Sampling the Yukawa couplings for $m_S = m_F = 5$ TeV from Eq.~(\ref{eq:YSLQ_CI}), we obtain couplings to the first generation quarks of order $10^5$, i.e., well above the perturbative limit. More generally, the perturbativity requirement on the Yukawa-like couplings would push the possible new physics scale far beyond the multi-TeV range in this setup, and one would never be able to observe $0\nu\beta\beta$.

\subsection{Two generations of $S_3$ }
Since the minimal choice with one generation of $S_3$ discussed in the previous subsection does not reproduce the observed neutrino masses and mixing patterns for perturbative Yukawa couplings, it is natural to extend the analysis to two generations of $S_3$, which we will refer to as $S_1$ and $S_2$ in this section. It is also easy to see that both $S_1$ and $S_2$ need to be coupled to second- and third-generation leptons to reproduce the correct neutrino masses. If one leptoquark generation, say $S_1$, were to couple only to electrons, it would only generate an effective first-generation neutrino mass $M_{ee}^{S_1}$ and we would have to fit the modified neutrino mass matrix $\hat{M}_{\nu\nu} = M_{\nu\nu} - \text{diag}(M_{ee}^{S_1}, 0, 0)$ with $S_2$. This task would lead to the same problem as in the previous section due to huge hierarchies in the loop factor.

Therefore, we conclude that both generations of $S_3$ are required to couple to all lepton generations to reproduce the correct neutrino masses. We follow the prescription described in~\cite{Angel:2013hla}: having two leptoquark generations, we can safely neglect all first- and second-generation quarks in the loop, which was not possible in the previous setup with a single leptoquark generation. In the following, $y_{SLQ}^{p3\alpha}$ denotes the coupling of the leptoquark generation $\alpha$ to the $p$th lepton generation and the third quark generation. Similarly, $Y_{SdF}^{3\alpha}$ denotes the coupling of the leptoquark generation $\alpha$ to the bottom quark and $F_8$.

\subsubsection{Neutrino masses}
Focusing on third-generation bottom quarks with mass $m_b$ in the loops of the 2-loop Weinberg diagram in Fig.~\ref{fig:M11_diagrams} (c), the neutrino mass matrix is given by
\begin{equation}
M_{\nu\nu}^{pt} =
     \sum_{\alpha, \beta=1}^{2} \left( y_{SLQ}^{p3\alpha}\right)^* I_{\alpha \beta} \left( y_{SLQ}^{t3\beta}\right)^* \,,
\end{equation}
where now $I$ is a $2 \times 2$ matrix with the entries $I_{\alpha\beta}$ given by
\begin{equation}
    I_{\alpha \beta} = 6\sqrt{2} m_{F_8} m_b^2 y_{SdF}^{3\alpha} y_{SdF}^{3\beta} \mathcal{I}_{bb,S_{\alpha}S_{\beta},F_8},
\end{equation}
and the integral $\mathcal{I}_{ab,\alpha \beta,X}$ as given in Eq.~(\ref{eq:I_definition}). The relevant mass ratios are in this case $r_b = \tfrac{m_b^2}{m_{F_8}^2}$ and $t_{\alpha} = \tfrac{m_{S_3^{\alpha}}^2}{m_{F_8}^2}$. In matrix form, we can write
\begin{equation}
M_{\nu\nu} = y_{SLQ}^* I y_{SLQ}^{\dagger},   
\end{equation}
where $y_{SLQ} = \left(y_{SLQ}^{p3\alpha}\right)_{p\alpha}$ is a $3\times 2$ matrix. 
Note that with two generations of leptoquarks and ${m_{S_1} \neq m_{S_2}}$, the loop factor $I$ is of rank-2, which is sufficient to give finite entries in the neutrino mass matrix and compatible with the current observation that the lightest neutrino can be massless. We diagonalise $I$ via a 2-dimensional unitary matrix $S$ and follow the same steps as detailed in the previous subsection.
The matrix $\mathcal{R}$ in Eq.\ (\ref{eq:R_definition}) is now a $2\times 3$ matrix
\begin{equation}
    \mathcal{R}_{2 \times 3} \equiv \sqrt{I_D} S^{\dagger} y_{SLQ}^{\dagger} U \sqrt{D_{\nu}^{-1}} \,,
\end{equation}
which can be written as
\begin{equation}
\mathcal{R}_{2 \times 3} = \begin{pmatrix}
    \mathbb{1} & | & 0_{2\times 1}
\end{pmatrix} \times \mathcal{R}_{3\times 3},
\end{equation}
with $\mathcal{R}_{3\times3}$ being the proper 3-dimensional orthogonal matrix given in Eq.~(\ref{eq:R_parametrisation}).
Solving for $y_{SLQ}$ gives 
\begin{equation}
    y_{SLQ} = U \sqrt{D_{\nu}} \mathcal{R}_{2\times3}^T \sqrt{\left(I_D^{-1}\right)^*} S^{\dagger}
    \label{eq:YSLQ_CI_2gen}.
\end{equation}

This parametrisation depends on the three free input parameters $\Lambda = m_{F_8}$, $t_1 = \frac{m_{S_1}^2}{\Lambda^2}$ and $t_2 = \frac{m_{S_2}^2}{\Lambda^2}$ and the five angles $\alpha, \beta, \Theta_a, \Theta_b$ and $\Theta_c$. For fixed $\Lambda$, $t_1$ and $t_2$, we sampled three thousand points for $y_{SLQ}$, using random values for the five angles. Note that for $ t_1 = t_2$, the loop factor $I_{\alpha \beta}$ would become degenerate and non-invertible, leading to the same problems as in the previous section for a single leptoquark generation. We checked that for $\Lambda$ in the few-TeV ballpark and $t_i < 1$, the sampled Yukawa couplings remain perturbative. 

Although the model discussed here in detail agrees with the model in~\cite{Angel:2013hla}, our analysis differs in one crucial point: Ref.~\cite{Angel:2013hla} focuses on the case in which all complex phases $\delta, \alpha, \beta$ and the angles $\Theta_a$, $\Theta_b$, $\Theta_c$ are set to zero and scans over different values for $t_1$ and $t_2$. In our analysis, we adapt benchmark scenarios for fixed values of $m_{F_8}$, $t_1$, and $t_2$ and sample solutions from the Casas-Ibarra parametrisation in Eq.~(\ref{eq:YSLQ_CI_2gen}) for random values of these angles.

Although for neutrino masses, and hence the Weinberg operator contribution to $0\nu\beta\beta$, it is sufficient to consider BSM Yukawa couplings to third-generation quarks, the tree-level dimension-9 operators require couplings to first-generation quarks. These couplings are not constrained by neutrino masses in this setup.
Therefore, to compare the $0\nu\beta\beta$ induced by dimension-9 and dimension-5 operators, we need to make further assumptions.

\subsubsection{$0\nu\beta\beta$ and CLFV for quark-flavour universal Yukawa coupling}
\label{sec:0vbb-Volkas}

Without further information about the BSM model, it is natural to assume that the leptoquarks couple to all quark flavours universally. Hence, first, we set $y_{SLQ}^{ip\alpha} = y_{SLQ}^{i3\alpha}$ for $p= 1,2$ for both leptoquark generations ($\alpha = 1,2$) and all lepton flavours ($i=1,2,3$). We study five benchmark scenarios for the scalar mass ratios $t_1$ and $t_2$, each for the two choices $m_F = 5$ TeV and $m_F = 10$ TeV:
\begin{align}
\label{eq:benchmark_CI}
    & \emph{B1}\text{: } t_1 = 0.49, t_2 = 0.51 \,, \qquad \emph{B2}\text{: } t_1 = 0.89, t_2 = 0.91\,, \qquad \emph{B3}\text{: } t_1 = 0.5, t_2 = 0.9  \,, \\
    & \emph{B4}\text{: } t_1 = 0.251, t_2 = 0.9 \,, \qquad \emph{B5}\text{: } t_1 = 0.251, t_2 = 0.5 \,. \nonumber
\end{align}

The limit $m_S > 2.5$ TeV from direct searches for leptoquarks puts a lower bound of 0.25 on $t_1$ and $t_2$ for $m_F = 5$ TeV. Note that for $t_i = 0.25$, the loop factor diverges, and that the case $t_1 = t_2$ is not possible because degenerate scalar masses cannot yield a rank-2 loop factor (similar to the situation with only one leptoquark discussed above). Benchmarks \emph{B1} and \emph{B2} focus on regions of the parameter space with $t_1 \approx t_2$, for which we will see that the $0\nu\beta\beta$ rate can be increased, while \emph{B3-B5} examine larger mass splittings between the scalars. As explained above, unlike Ref.~\cite{Angel:2013hla}, we allow for $CP$ violating phases $\delta$, $\alpha$ and $\beta$ and a non-trivial mixing matrix $\mathcal{R}_{2 \times 3}$.
Again, we sample $y_{SLQ}$ in Eq.~\eqref{eq:YSLQ_CI_2gen} for three thousand random angle choices for each of the five benchmark scenarios.

\begin{figure}
    \centering
    \begin{subfigure}[t]{0.49\textwidth}
         \includegraphics[width=\linewidth]{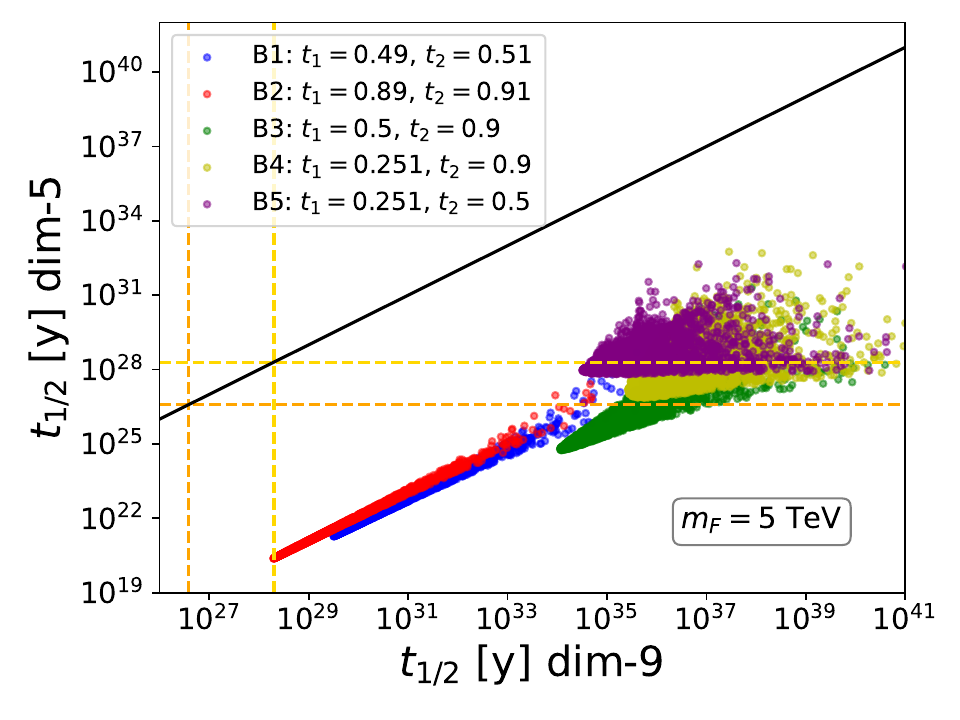}
    \end{subfigure}
      \begin{subfigure}[t]{0.49\textwidth}
         \includegraphics[width=\linewidth]{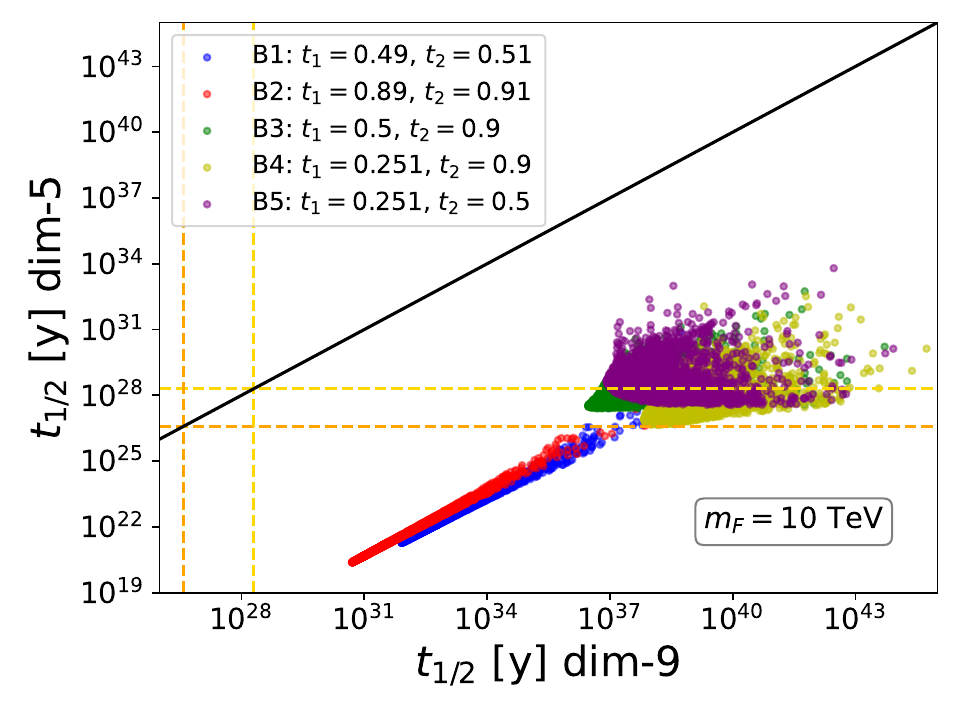}
    \end{subfigure}
    \caption{We depict the $0\nu\beta\beta$ half-life in $^{136}$Xe induced by dimension-9 ($x$-axis) and dimension-5 ($y$-axis) operators, respectively. The points in each colour show the half-lives for different random samples for $y_{SLQ}$ from Eq.~(\ref{eq:YSLQ_CI_2gen}) for the five benchmark scenarios in Eq.~(\ref{eq:benchmark_CI}). Along the black solid line, the contributions from dimension-9 and dimension-5 operators are equal. The orange dashed lines show the current exclusion limit from KamLAND-Zen ($3.8 \cdot 10^{26}$ years) and the yellow dashed lines the expected future sensitivity ($2 \cdot 10^{28}$ years).  }
    \label{fig:half_lives_Volkas}
\end{figure}

\paragraph{Neutrinoless double beta decay:} We calculate the expected $0\nu\beta\beta$ half-life triggered by each class of operators for all the samples with \texttt{$\nu$DoBe}. Fig.~\ref{fig:half_lives_Volkas} compares the half-lives that originate from the dimension-9 and dimension-5 operators, respectively, for $m_F =5$ TeV (left) and $m_F = 10$ TeV (right). The different colours represent the five different benchmark models in Eq.~(\ref{eq:benchmark_CI}). For all points below the black solid line, the dimension-5 half-life is shorter than the dimension-9 half-life, and hence the Weinberg operator dominates the $0\nu\beta\beta$ for all benchmarks. Estimating individual half-lives showed that the dimension-7 contributions are always subdominant compared to the dimension-9 and/or dimension-5 contributions, so that we decided to plot the dimension-9 versus the dimension-5 contributions to $0\nu\beta\beta$. The orange and yellow dashed lines show the current KamLAND-Zen limit and the projected nEXO sensitivity, respectively. The points between these two lines would correspond to appealing models that are not excluded by current experiments and are testable in next-generation searches. Although we find points with dimension-5-induced half-lives in this band, the dimension-9 half-lives are rather large. As expected, the shortest half-lives come from the benchmarks for which $t_1 \sim t_2$. 

We mention in passing that the Yukawa couplings $y_{Sue}$, which couple the leptoquarks to quark and electron singlets and enter the dimension-9 operator matching, are not constrained by neutrino masses. However, the half-life induced by the operator $\mathcal{O}^{(9)}_{ddueue}$ and triggered by $y_{Sue}$, is of the order of $10^{32}$ years for $y_{Sue}^{11} = 1$ and $\Lambda = 5$ TeV. Thus, it is several orders of magnitude above the $y_{SLQ}$ contribution and also above the future experimental sensitivity. In this work, we focus exclusively on the couplings $y_{SLQ}$, which can, in principle, lead to observable $0\nu\beta\beta$.

Also note that we set all unknown low-energy constants (LECs) to zero in the calculations that lead to Fig.~\ref{fig:half_lives_Volkas}. Switching on the unknown LECs, we can lower the half-lives by several orders of magnitude, as we confirm explicitly in Appendix~\ref{sec:LECs}. However, the dimension-5 operator will still dominate. 

\paragraph{Charged lepton flavour violation:} By its nature, $0\nu\beta\beta$ probes the couplings of the leptoquarks to the first-generation leptons. To generate non-hierarchical neutrino masses, we obtain couplings of the same order of magnitude to all lepton flavours from the Casas-Ibarra parametrisation. This also allows the models to be tested with experimental probes of charged lepton flavour violation. For a detailed analysis of various possible constraints we point to~\cite{Angel:2013hla} and focus on two of the most promising probes here: $\mu \rightarrow e \gamma$ and $\mu$ to $e$ conversion in nuclei, which both test couplings to first and second lepton generations. While $\mu \to e \gamma$ is loop-induced in our model, the $\mu \to e$ conversion also receives tree-level contributions. In this work, we focus exclusively on the couplings $y_{SLQ}$ which can, in principle, lead to observable $0\nu\beta\beta$ and which are constrained by the neutrino masses. In the second step, we will allow for quark-flavour-universal couplings, in accordance with the assumption made previously for $0\nu\beta\beta$. 

\begin{figure}[t!]
    \centering
    \begin{subfigure}[t]{\textwidth}
    \centering
    \includegraphics[scale=0.9]{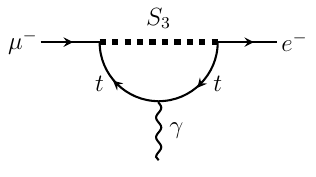}
    %\hskip5mm
    \includegraphics[scale=0.9]{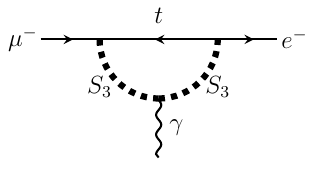}
    \caption{Diagrams for $\mu \rightarrow e \gamma$}
    \end{subfigure}
    \begin{subfigure}[t]{\textwidth}
    \centering
    \includegraphics[scale=0.9]{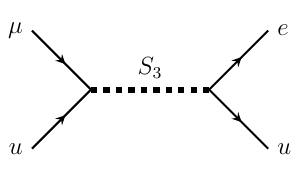}
    \caption{Tree-level diagram for $\mu$ to $e$ conversion}
    \end{subfigure}
    \begin{subfigure}[t]{\textwidth}
    \centering
    \includegraphics[scale=0.9]{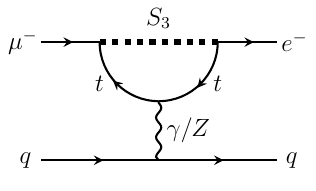}
    %\hskip5mm
    \includegraphics[scale=0.9]{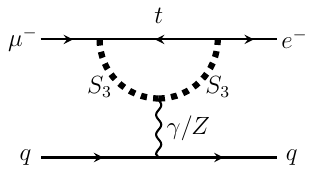}
    \caption{Penguin diagrams for $\mu$ to $e$ conversion}
    \end{subfigure}
    \begin{subfigure}[t]{\textwidth}
    \centering
    \includegraphics[scale=0.9]{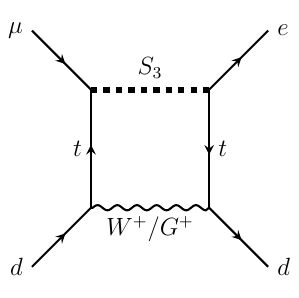}
    \includegraphics[scale=0.9]{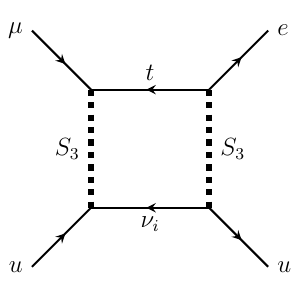}
    \caption{Box diagrams for $\mu$ to $e$ conversion.}
    \end{subfigure}
    \caption{Diagrams for $\mu \rightarrow e \gamma$ and $\mu \to e$ conversion in nuclei mediated by the leptoquark $S_3$.}
    \label{fig:CLFV_diagrams}
\end{figure}

The dominant contribution to $\mu \rightarrow e \gamma$ conversion is induced by the dipole operator. The corresponding diagrams are shown in Fig.~\ref{fig:CLFV_diagrams} (a). Note that the photon could also be radiated off the initial- or final-state lepton.
Assuming now only leptoquark couplings to the third-generation quarks and the final-state electron mass to be negligible, the branching ratio is given by~\cite{Angel:2013hla}
\begin{equation}
    \text{BR} (\mu \rightarrow e \gamma) \simeq \frac{3 s \theta_W^2 }{8 \pi^3 \alpha} |V_{tb}|^2 \sum_{\alpha=1}^{2} \left(y_{SLQ}^{23\alpha} \right)^* y_{SLQ}^{13\alpha} \frac{m_W^2}{m_{S_3}^2} F \left( t_{3\alpha} \right)\,,
\label{eq:BRmuegamma}
\end{equation}
with $t_{3\alpha} = \tfrac{m_t^2}{m_{S_{\alpha}}^2}$, the weak mixing angle $\theta_W$, $s \theta_W = \sin \theta_W$, and the fine structure constant $\alpha$. Since we work in the basis in which the down-type quark Yukawa couplings are diagonal in the mass basis, the up-type quark couplings will be proportional to the respective CKM elements, i.e., $V_{tb}$ in this case.
The loop function is given by
\begin{equation}
    F(t) = \frac{1+ 4 t - 5t^2 + 2 t(2+t) \ln(t)}{4 (t-1)^4}.
\end{equation}
We notice that for small $t_{3\alpha}$, i.e., for the new physics scale heavy compared to all SM quark masses, $F(t)$ goes asymptotically to $\frac{1}{4}$.

The $\mu \to e$ conversion rate in nuclei for this model reads
\begin{equation}
    \omega_{\rm conv} = 4 \Big| \frac{1}{8} A_2^{R*} D + \tilde{g}_{LV}^{(p)} V^{(p)} + \tilde{g}_{LV}^{(n)} V^{(n)} \Big|^2 + 4 \Big| \frac{1}{8} A_2^{L*} D\Big| ^2,
\label{eq:omega_conv}
\end{equation}
where $D$, $V^{(p)}$ and $V^{(n)}$ are isotope-dependent constants. For the $_{79}^{197}$Au isotope, one finds $D=0.108 \left(m_{\mu}\right)^{\tfrac{5}{2}}$, $V^{(p)}=0.0610 \left(m_{\mu}\right)^{\tfrac{5}{2}}$ and $V^{(n)} = 0.0859 \left(m_{\mu}\right)^{\tfrac{5}{2}}$ \cite{Kitano:2002mt}. The branching ratio (BR) is defined by
\begin{equation}
    \text{BR}(\mu N \rightarrow e N) = \frac{\omega_{\rm conv}}{\omega_{\rm capt}},
\end{equation}
where $\omega_{\rm capt}$ denotes the total capture rate and $\omega_{\rm capt}(\rm Au) = 13.07 \cdot 10^6 s^{-1}$. In Eq.~(\ref{eq:omega_conv}),
\begin{equation}
    A_2^L = \frac{m_e}{16 \pi^2 m_{\mu}} |V_{tb}|^2\sum_{\alpha=1}^2 \frac{\left(y_{SLQ}^{23\alpha}\right)^* y_{SLQ}^{13\alpha}}{m_{S_{\alpha}}^2} F(t_{3\alpha}),
\end{equation}
and 
\begin{equation}
    A_2^R = \frac{1}{16 \pi^2} |V_{tb}|^2\sum_{\alpha=1}^2 \frac{\left(y_{SLQ}^{23\alpha}\right)^* y_{SLQ}^{13\alpha}}{m_{S_{\alpha}}^2} F(t_{3\alpha})
\end{equation}
denote the long-range contributions. The short-range vector part
\begin{equation}
\begin{split}
    \tilde{g}_{LV}^{(p)} & = 2 g_{LV(u)} + g_{LV(d)} \,, \\
    \tilde{g}_{LV}^{(n)} & =  g_{LV(u)} + 2 g_{LV(d)}\,,  
\end{split}
\end{equation}
originates from interactions with up and down quarks in the nucleus,
\begin{equation}
\begin{split}
    g_{LV(u)} &= g_{LV(u)}^{box} + g_{LV(u)}^{\gamma} + g_{LV(u)}^{Z} + g_{LV(u)}^{tree}\,, \\
    g_{LV(d)} &= g_{LV(d)}^{box} + g_{LV(d)}^{\gamma} + g_{LV(d)}^{Z} \,.
\end{split}
\end{equation}
The individual contributions from box diagrams, $\gamma$- and Z-penguins for the down quarks, cf.\ Fig.\ \ref{fig:CLFV_diagrams} (d) and (c), respectively, are given by
\begin{equation}
\begin{split}
    g_{LV(d)}^{box} & = - \frac{\big|V_{tb} \big|^2 \big|V_{td} \big|^2}{64 \pi^2}\sum_{\alpha} \left(y_{SLQ}^{23\alpha}\right)^* y_{SLQ}^{13\alpha} \Big\{ m_t^2 y_t^2 D_0 \left( m_W^2, m_{S_{\alpha}}^2, m_t^2, m_t^2 \right) \\
    & + g^2 \left( C_0 \left(m_W^2,0,m_{S_{\alpha}}^2 \right) + m_t^2 D_0 \left(m_W^2,0,m_{S_{\alpha}}^2, m_t^2\right) \right.  \\
    & - \left. 2 \left( D_{00} \left(m_W^2,0,m_{S_{\alpha}}^2,m_t^2\right) +D_{00} \left( m_W^2,m_{S_{\alpha}}^2, m_t^2, m_t^2 \right) \right) 
    \right)  \Big\} \,, \\
    g_{LV(d)}^{\gamma} &= -\frac{\alpha \big|V_{tb} \big|^2}{144 \pi} \sum_{\alpha} \frac{\left(y_{SLQ}^{23\alpha}\right)^* y_{SLQ}^{13\alpha}}{m_{S_{\alpha}}^2} \frac{t_{3\alpha}^3 - 18 t_{3\alpha}^2 + 27 t_{3\alpha} + 2\left(t_{3\alpha}^3 + 6 t_{3\alpha} -4 \right) \ln \left(t_{3\alpha}\right) - 10}{\left(t_{3\alpha}-1\right)^4} \,, \\
    g_{LV(d)}^{(Z)} &= \frac{g^2\left(4 s_w^2-3\right)\big|V_{tb} \big|^2}{128 \pi^2 m_W^2} \sum_{\alpha} \left(y_{SLQ}^{23\alpha}\right)^* y_{SLQ}^{13\alpha} \frac{t_{3\alpha}\left(t_{3\alpha}-\ln\left(t_{3\alpha}\right) -1 \right)}{\left(t_{3\alpha}-1\right)^2}\,.
\end{split}
\end{equation}

For up-quarks, the leptoquark can mediate $\mu \to e$ conversion at tree level, see Fig.\ \ref{fig:CLFV_diagrams} (b),
\begin{equation}
    g_{LV(u)}^{tree} = - \big| V_{ub}\big|^2\sum_{\alpha}\frac{1}{m_{S_{\alpha}}^2}  \left( Y_{SLQ}^{23\alpha} \right)^* Y_{SLQ}^{13\alpha} \,.
\end{equation}
At the 1-loop level, one finds box and penguin contributions too, with the latter ones related to the down-quarks,
\begin{equation}
    \begin{split}
    g_{LV(u)}^{box} 
     & = \frac{\big|V_{tb} \big|^2 \big|V_{ub} \big|^2}{64 \pi^2}  2 \sum_{\alpha,\beta,i} \left(y_{SLQ}^{23\alpha}\right)^* y_{SLQ}^{13\beta} \left(y_{SLQ}^{i3\beta}\right)^* y_{SLQ}^{i3\alpha} D_{00}\left( 0, m_{S_{\alpha}}^2, m_{S_{\beta}}^2, m_t^2\right) \,,\\
    g_{LV(u)}^{\gamma} &= - 2 g_{LV(d)}^{\gamma}\,,\\
    \quad g_{LV(u)}^Z &= -\frac{8 s_W^2 -3}{4 s_W^2-3} g_{LV(d)}^Z \,.
\end{split}
\end{equation}
The corresponding penguin diagrams are shown in Fig.~\ref{fig:CLFV_diagrams} (b).
Note that if we consider only leptoquark couplings to third generation quarks, the box diagrams in Fig.~\ref{fig:CLFV_diagrams} (c) contribute only through CKM mixing in both cases, $\propto \big| V_{td} \big|^2$ for down quarks and $\propto \big| V_{ub} \big|^2$ for up quarks. The same holds for the tree-level diagram that is suppressed by $\big|V_{ub}\big|^2$. In addition, note that choosing the Yukawa alignment of the down-type, the individual box contributions differ from~\cite{Angel:2013hla}, but agree with the convention in~\cite{Dorsner:2016wpm}.

\begin{figure}
    \centering
    \begin{subfigure}[t]{0.49\textwidth}
         \includegraphics[width=\linewidth]{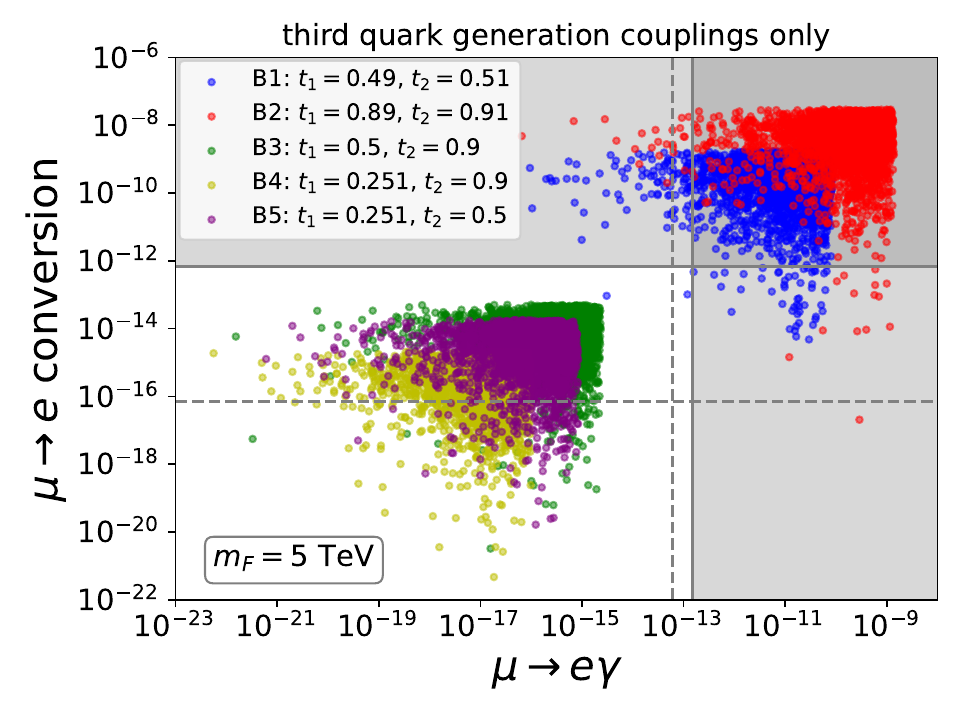}
    \end{subfigure}
      \begin{subfigure}[t]{0.49\textwidth}
        \includegraphics[width=\linewidth]{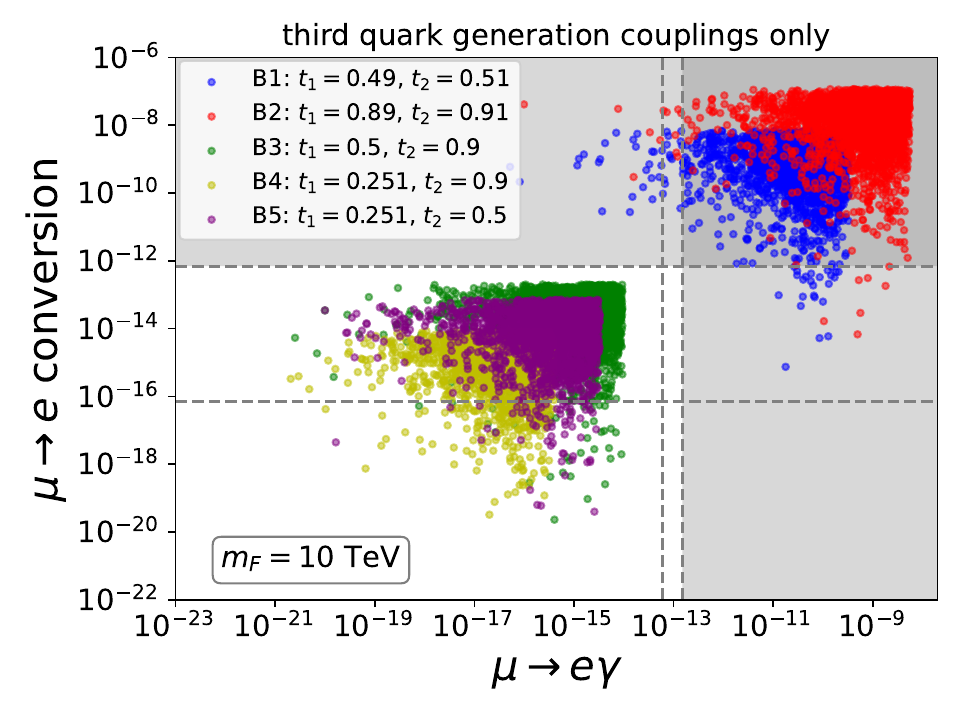}
    \end{subfigure}
    \caption{Branching ratios for $\mu\to e\gamma$ and $\mu \to e $ conversion in gold nuclei for $m_F = 5$~TeV (left) and $m_F = 10$~TeV (right) for the benchmark scenarios in Eq.\ (\ref{eq:benchmark_CI}), assuming leptoquark couplings to third generation quarks only. The grey shaded areas are excluded experimentally, and the dashed grey lines show the expected future sensitivity.}
    \label{fig:CLFV_Volkas}
\end{figure}

Figure~\ref{fig:CLFV_Volkas} shows the branching ratios for $\mu \to e \gamma$ versus $\mu \to e$ conversion in gold nuclei for the 5 benchmark scenarios in Eq.~(\ref{eq:benchmark_CI}) for $m_F = 5$ TeV (left) and $m_F = 10$ TeV (right). The gray shaded areas are excluded by the current experimental limits $\text{BR}(\mu \to e \gamma) < 5.7 \cdot 10^{-13}$ \cite{MEG:2013oxv} and $\text{BR}(\mu \,\text{Au} \to e \,\text{Au}) < 7 \cdot 10^{-13}$ \cite{PhysRevD.86.010001}, and the dashed gray lines show the predicted future sensitivities, $\text{BR}(\mu \to e \gamma) \lesssim 6 \cdot 10^{-14}$ \cite{MEGII:2018kmf} and $\text{BR} (\mu~\to~e )~\lesssim~7~\cdot~10^{-17}$~\cite{COMET:2018wbw}. Points in the same colour correspond to samples from the Casas-Ibarra parametrisation for the same benchmark scenario and different values of the angles $\alpha, \beta, \Theta_1, \Theta_2$ and $\Theta_3$. For benchmarks with small mass splitting, \emph{B1} and \emph{B2} in blue and red, respectively, the predicted branching ratios are larger than for benchmarks with larger mass splitting. Although these benchmarks could enhance the $0\nu\beta\beta$ rate as seen in Fig.~\ref{fig:half_lives_Volkas}, they are excluded by both $\mu \to e \gamma$ and $\mu \to e$ conversion. The models with larger scalar mass splitting, \emph{B3}, \emph{B4} and \emph{B5}, on the other hand, are compatible with the experimental cLFV limits, but result in large $0\nu\beta\beta$ half-lives induced by dimension-9 operators. 

\begin{figure}[t]
    \centering
    \begin{subfigure}[t]{0.49\textwidth}
         \includegraphics[width=\linewidth]{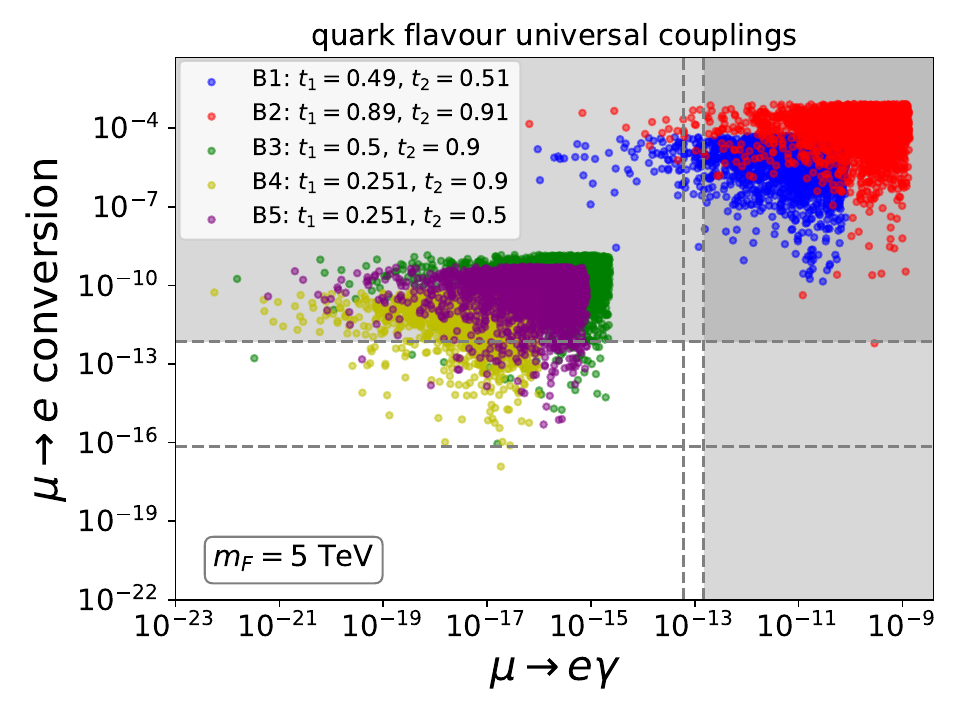}
    \end{subfigure}
      \begin{subfigure}[t]{0.49\textwidth}
        \includegraphics[width=\linewidth]{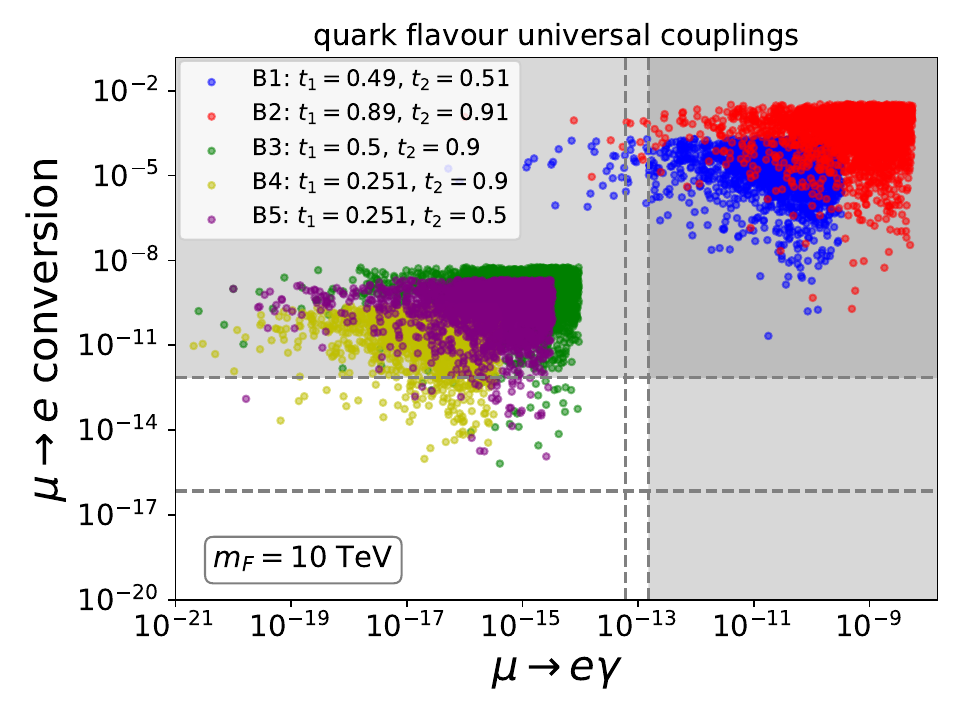}
    \end{subfigure}
    \caption{Branching ratios for $\mu\to e\gamma$ and $\mu \to e $ conversion in gold for $m_F = 5$~TeV (left) and $m_F = 10$~TeV (right) for the benchmark scenarios in \ref{eq:benchmark_CI}, assuming quark-flavour universal leptoquark couplings. The grey shaded areas are excluded experimentally, and the dashed grey lines show the expected future sensitivity. }
    \label{fig:CLFV_Volkas_FU}
\end{figure}

Next, assuming that the leptoquarks couple flavour universally to all quark generations, the first- and second-generation quarks will also contribute to $\mu \to e $ conversion. In this case, all vertices with up quarks coupling to the leptoquark are proportional to $\sum_{k=1}^{3} y_{SLQ}^{ik\alpha} V_{1k}$ and with down quarks proportional to $\sum_{k=1}^{3} y_{SLQ}^{ik\alpha} V_{k1}$. For quark-flavour universal $y_{SLQ}$, the dominant contribution will arise from $y_{SLQ}^{i1\alpha} V_{ud}$ in both cases. In contrast to the calculations in the third-generation-only case above, tree-level and box-diagrams are not CKM suppressed, and the prediction for $\mu \to e$ conversion is enhanced with respect to the previous calculation by a factor of $\frac{\big|V_{ud}\big|^2}{\big|V_{ub}\big|^2}$ for up quarks and $\frac{\big|V_{ud}\big|^2}{\big|V_{td}\big|^2}$ for down quarks, which both amount to roughly four orders of magnitude. The prediction for $\mu \to e \gamma$ is not affected by this assumption. We show the resulting predictions in Fig.~\ref{fig:CLFV_Volkas_FU}. As expected, the $\mu \to e$ conversion rate is roughly four orders of magnitude larger than in Fig.~\ref{fig:CLFV_Volkas}. This implies that even for the benchmark scenarios \emph{B3}, \emph{B4} and \emph{B5} with larger scalar mass splittings, the bulk of the parameter space is excluded by the current upper limit on $\mu \to e$ conversion. In summary, assuming quark-flavour universal leptoquark couplings in this model can not simultaneously reproduce the correct neutrino masses, be compatible with cLFV constraints and lead to a dominant dimension-9 contribution to $0\nu\beta\beta$.

\subsubsection{$0\nu\beta\beta$ for quark-flavour non-universal Yukawa coupling}

\begin{figure}[t]
    \centering
\includegraphics[scale=0.6]{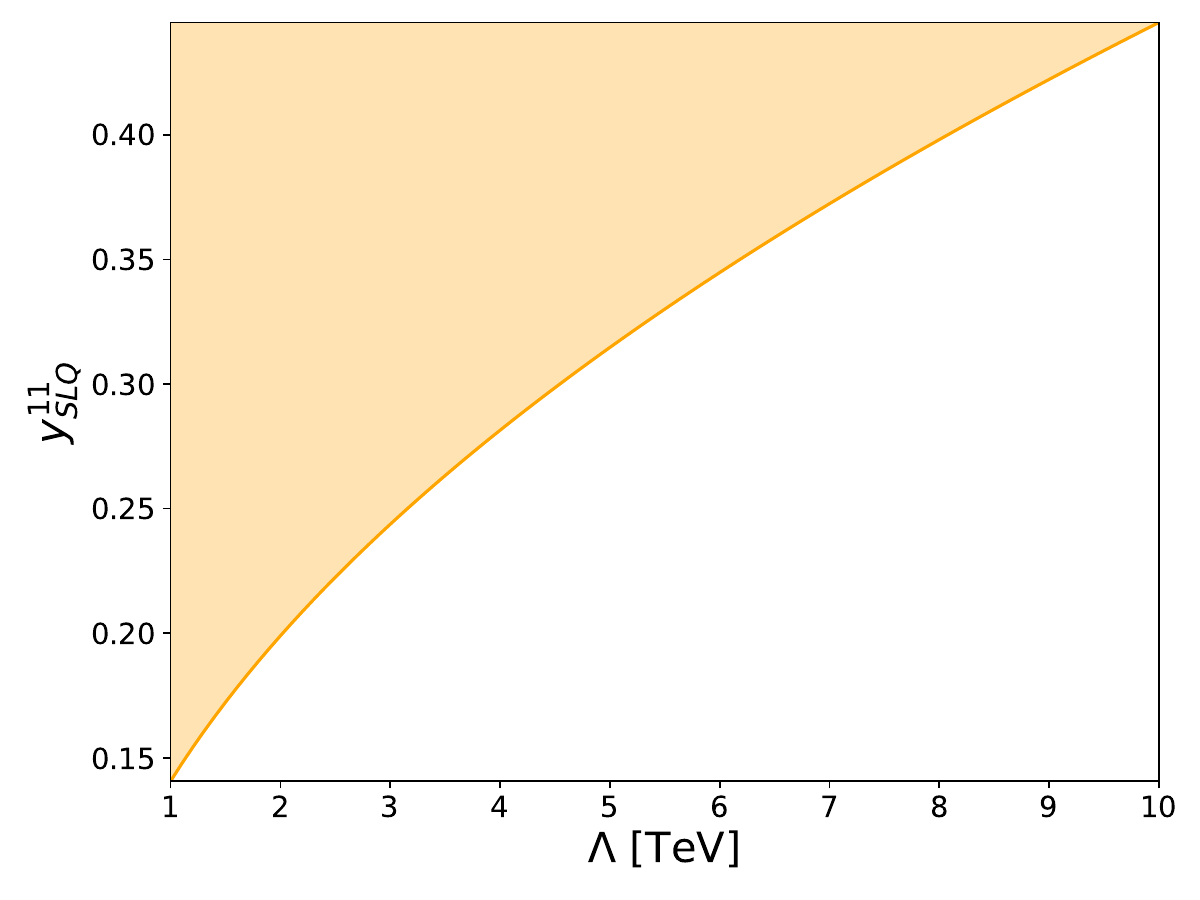}
    \caption{Allowed values for $y_{SLQ}^{11}$ as function of $\Lambda$. All points on the orange line give $m_{\beta\beta} = 10^{-3}$ eV, meaning that values of $y_{SLQ}^{11}$ below it are compatible with the assumption $m_{\beta\beta} < 10^{-3}$ and all values in the orange shaded area are excluded.}
\label{fig:M3_YSLQ_Mbb}
\end{figure}

As a second option, we exploit a quark-flavour non-universal scenario, which is motivated by enhancing the rate for $0\nu\beta\beta$ induced by dimension-9 operators without spoiling the generation of the neutrino masses and without clashing with cLFV bounds. To this end, we make use of the following observations:
\begin{enumerate}
    \item[(i)] The tree-level contributions of dimension-9 operators to $0\nu\beta\beta$ depend only on $y_{SLQ}^{11\alpha}$.
    \item[(ii)] We saw in the previous section that a model with two leptoquark generations that couple only to third generation quarks and a sufficient large scalar mass splitting (\emph{B3}, \emph{B4}, \emph{B5} in Eq.~(\ref{eq:benchmark_CI})) can reproduce the neutrino masses and evade cLFV constraints.
    \item[(iii)] Both $\mu \to e \gamma$ and $\mu \to e$ conversion require leptoquark couplings to both muons and electrons. Switching on the couplings to electrons and keeping the couplings to muons zero will thus not add any new contributions to either of the two cLFV processes. 
\end{enumerate}
Hence, our requirements can be fulfilled for $y_{SLQ}^{11\alpha}$ as large as possible without significantly affecting neutrino masses. 
For this consideration, it is not important to have two active leptoquark generations coupling to first-generation quarks as long as the contributions from both to the neutrino mass are small enough. In fact, we only need to know about the two generations that $y_{SLQ}^{p3\alpha}$ can give the correct neutrino masses as discussed above. For first-generation couplings, we will now return to considering only one leptoquark generation with mass $m_{S_1}=\Lambda$. We saw above that it is always possible to choose the mass of the other leptoquark generation, $m_{S_2}$ and the couplings to the third-generation quarks such that the neutrino masses are reproduced, and the model remains compatible with cLFV constraints.
We denote these first-generation couplings by $y_{SLQ}^{11}$. Then the statement can be quantified by imposing that the effective electron-neutrino mass induced by $y_{SLQ}^{11}$ satisfies {$m_{\beta\beta}\left(y_{SLQ}^{11}\right) < 10^{-3}$ eV}.
This gives a relation between the values for $y_{SLQ}^{11}$ that are allowed for a given scale $\Lambda$, which is shown in Fig.~\ref{fig:M3_YSLQ_Mbb}. All points in the shaded area would give a too large contribution to $m_{\beta\beta}$ and are excluded, while all points on the orange line give precisely $m_{\beta\beta} = 10^{-3}$ eV and all points below that curve affect the neutrino masses even less. Since we want to maximise the $0\nu\beta\beta$ rate, we choose for each $\Lambda$ the value of $y_{SLQ}^{11}$ with $m_{\beta\beta} \left(y_{SLQ}^{11} \right) = 10^{-3}$.

\begin{figure}[t!]
    \centering
\includegraphics[scale=0.6]{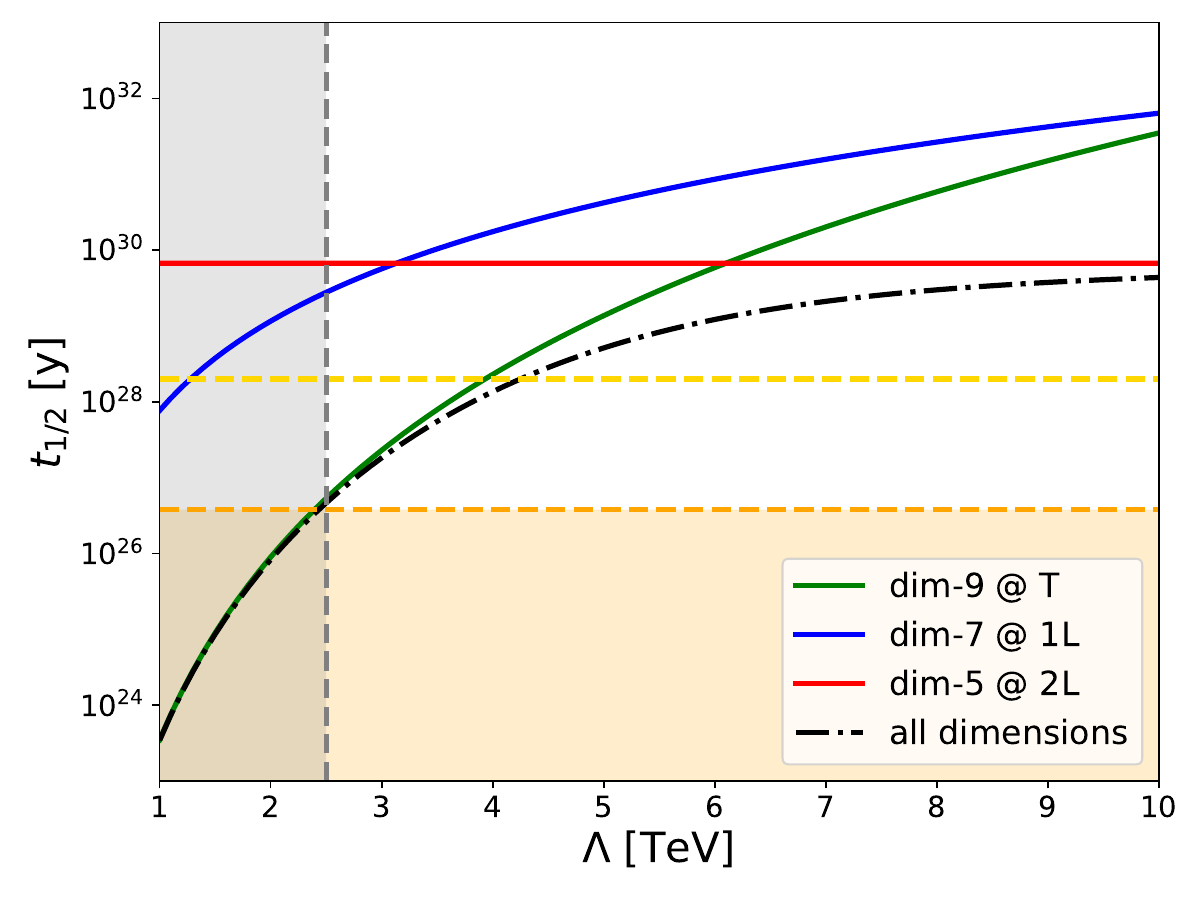}
    \caption{We depict the half-life of $0\nu\beta\beta$ in $^{136}$Xe induced by dimension-9 (green), dimension-7 (blue) and dimension-5 (red) operators as function of $\Lambda$, while fixing $Y_{SLQ}^{11}$ such that $m_{\beta\beta} = 10^{-3}$ eV. The black dashed-dotted line shows the combined contribution of all operators at a time, the orange dotted line represents the current exclusion limit from KamLAND-Zen ($3.8 \cdot 10^{26}$ years) and the yellow dashed line is the projected nEXO sensitivity ($2 \cdot 10^{28}$ years). Masses below $2.5$~TeV (shaded in grey) are excluded from direct searches for leptoquarks.}
\label{fig:M3_dim9vsdim5vsdim5_Mbb}
\end{figure}

We calculate the $0\nu\beta\beta$ contributions from the dimension-9, dimension-7 and dimension-5 operators for this fixed value of $y_{SLQ}^{11}(\Lambda)$ and we show them in Fig.~\ref{fig:M3_dim9vsdim5vsdim5_Mbb} in green, blue and red, respectively. The black dashed-dotted curve shows the combined contribution. We choose $y_{Sue}=0$, as we have already seen before, because it does not affect the half-life of $0\nu\beta\beta$ as long as $y_{SLQ}$ is large enough.
The Weinberg operator contributes to $0\nu\beta\beta$ directly through $m_{\beta\beta}$, so fixing the Yukawa coupling as a function of $\Lambda$ such that $m_{\beta\beta}$ is constant naturally implies that the dimension-5-induced half-life as a function of $\Lambda$ is also constant. As before, the gray area below $2.5$~TeV is excluded by direct searches for leptoquarks. The orange-shaded area is excluded from the observed lifetime in KamLAND-Zen. For our model to be testable, the prediction for the half-life should be smaller than the future sensitivity of $2 \times 10^{28}$ years, which is depicted by the yellow dashed line. 

We see that for $\Lambda < 5$~TeV, the dimension-9 contributions will dominate over the dimension-5 contributions. For $\Lambda$ between $2.5$~TeV and $4.2$~TeV we are furthermore in the most appealing range, which is not excluded by current observations, but will be probed in the next generation of experiments.
For example, $m_{S_3} = 4$ TeV and $y_{SLQ}^{11} = 0.28$ yield an interesting model that satisfies all the previously stated criteria.

\section{Conclusions}
\label{sec:conclusion}
Beyond the standard light-neutrino exchange mechanism, neutrinoless double beta decay ($0\nu\beta\beta$) can be mediated by a broad set of $\Delta L=2$ operators in the SMEFT, which arise at odd operator dimensions. In this work we focused on the dimension-9 operators that can contribute to $0\nu\beta\beta$ already at tree level. We systematically analysed their tree-level ultraviolet completions and identified the particularly interesting situations in which the associated dimension-7 and dimension-5 contributions are generated only radiatively. We then investigated representative model classes in the light of $0\nu\beta\beta$, radiative neutrino-mass generation and its compatibility with oscillation data, as well as current bounds from charged-lepton flavour violation (cLFV).

We developed a diagrammatic, step-by-step classification strategy: for a given operator and loop order, we first enumerate all relevant topologies, then promote them to renormalisable diagrams by inserting heavy mediators, and finally determine the corresponding models by assigning gauge quantum numbers. A central requirement in our construction of tree-level completions of the dimension-9 operators relevant for $0\nu\beta\beta$ was the absence of any tree-level generation of the Weinberg operator or of $\Delta L=2$ operators at dimension 7. Imposing this criterion yields 315 viable completions. Among them, 14 models introduce only two distinct BSM multiplets and no new vectors; we refer to these as \emph{minimal models}. They contain either two BSM scalars or one BSM scalar together with one vector-like fermion. We further organised these minimal scenarios by (i) the loop order at which they induce the Weinberg and dimension-7 operators and (ii) whether they can lead to symmetric neutrino-mass matrices.

As a concrete illustration, we analysed in detail a model containing a scalar leptoquark $S_{8,2,\frac{1}{2}}$ and a vector-like fermion $F_{8,1,0}$, which generates the Weinberg operator at one loop. We matched the model onto the relevant SMEFT operators and showed that, when scalar quartic couplings are symmetry-protected, the short-range dimension-9 contribution can dominate the $0\nu\beta\beta$ amplitude. Under more general assumptions, however, the radiatively induced Weinberg operator is not sufficiently suppressed and tends to dominate.

We also derived a general expression for the neutrino-mass matrix in the class of models that generate a symmetric Weinberg operator at two loops, finding that different realisations mainly differ by group-theory factors of order $\mathcal{O}(1\text{--}10)$. Within this class, we studied a specific model with a scalar leptoquark $S_{3,1,-\frac{1}{3}}$ and a fermionic colour octet $F_{8,1,0}$. Using the Casas--Ibarra parametrisation, we sampled couplings that reproduce neutrino oscillation data and found that two generations of the leptoquark are required to keep the couplings perturbative. In this setup, neutrino-mass constraints primarily restrict couplings involving third-generation quarks. We then examined the parameter space consistent with cLFV limits, in particular $\mu\to e\gamma$ and $\mu\to e$ conversion in nuclei. Assuming quark-flavour-universal couplings, the dimension-5 contribution typically dominates over the dimension-9 one. By contrast, if couplings to first-generation quarks are suppressed and their impact on neutrino masses is correspondingly reduced, an interesting few-TeV mass window emerges in which the dimension-9 operators can dominate $0\nu\beta\beta$ and remain testable in forthcoming experiments.

Overall, our analysis of two-loop Weinberg scenarios shows that with appropriate flavour structure, one can generically achieve a dimension-9-dominated $0\nu\beta\beta$ rate, whereas for flavour-universal couplings, the two-loop suppression of the Weinberg operator is usually insufficient to offset the stronger EFT suppression of dimension-9 effects by the new-physics scale. On the other hand, models in which the Weinberg operator first appears only at $\geq 4$ loops fail to reproduce neutrino masses of the observed size. This points to models with a three-loop-suppressed Weinberg operator as a particularly promising ``sweet spot'' for scenarios in which $0\nu\beta\beta$ is predominantly driven by short-range, dimension-9 dynamics.

\section*{Acknowledgments}
We thank Martin Hirsch and Ricardo Cepedello for many useful discussions and for their help with the implementation of the code for the diagram-based approach used in this work. F.~E. thanks Javier Fuentes-Mart\'{i}n, Julie Pag\`{e}s and Anders Eller Thomsen for help with the package \texttt{Matchete}. F.~E is supported by Charles University through the project PRIMUS/24/SCI/013. L.~G. acknowledges support from the Dutch Research Council (NWO) under project number VI.Veni.222.318 and from Charles University through the project number
PRIMUS/24/SCI/013. C.~H. is funded by the Generalitat Valenciana under Plan Gen-T via CDEIGENT grant No. CIDEIG/2022/16. C.~H. would also like to acknowledge support from the Spanish grants PID2023-147306NB-I00 and CEX2023-001292-S (MCIU/AEI/10.13039/501100011033).\\

\newpage
\appendix
\section{Effect of the unknown low-energy constants on $0\nu\beta\beta$}
\label{sec:LECs}

\begin{figure}[h]
    \centering
    \begin{subfigure}[t]{0.49\textwidth}
         \includegraphics[width=\linewidth]{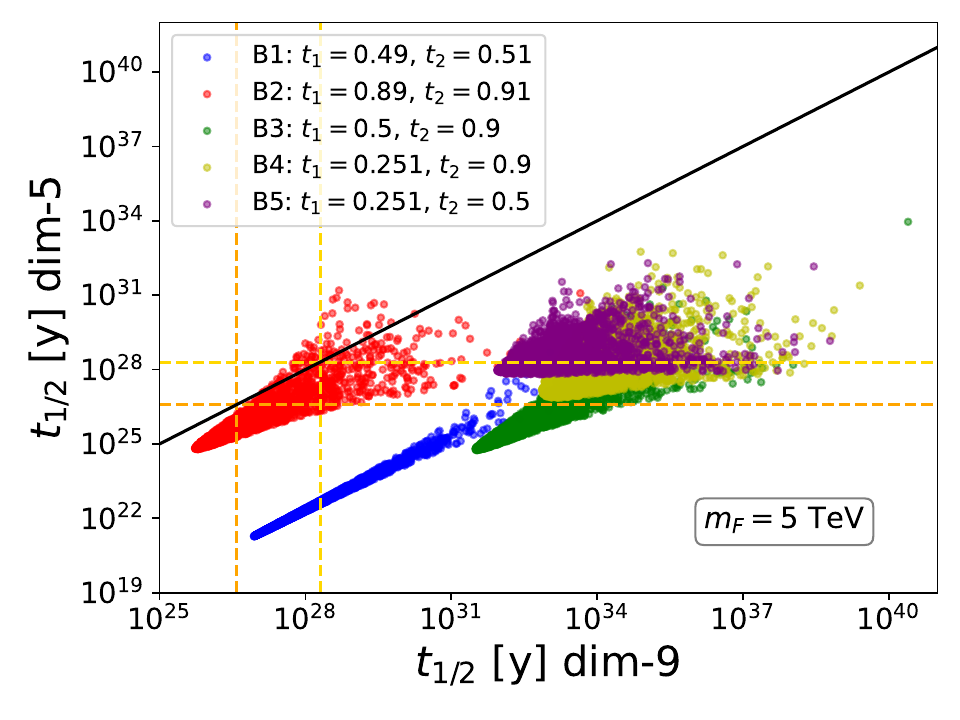}
    \end{subfigure}
      \begin{subfigure}[t]{0.49\textwidth}
         \includegraphics[width=\linewidth]{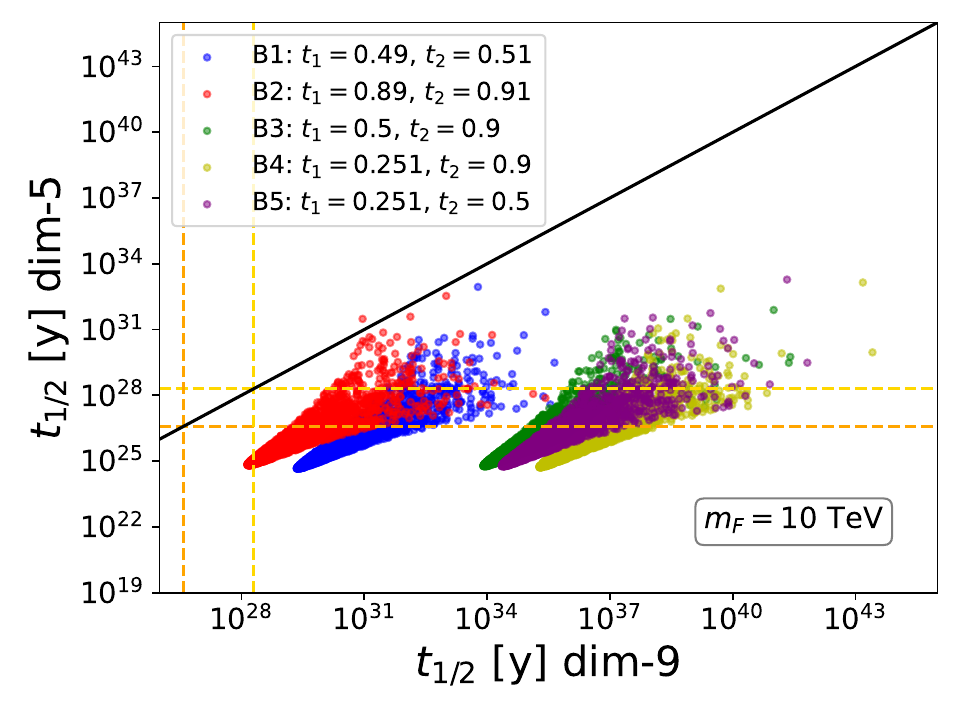}
    \end{subfigure}
    \caption{The half-life of $0\nu\beta\beta$ in $^{136}$Xe induced by dimension-9 (x-axis) and dimension-5 (y-axis) operators, respectively, obtained with setting the unknown LECs to their order-of-magnitude estimates in the \texttt{$\nu$DoBe} code. The points in each colour show the half-lives for different random samples for $y_{SLQ}$ from Eq.~(\ref{eq:YSLQ_CI_2gen}) for the five benchmark scenarios in Eq.~(\ref{eq:benchmark_CI}). Along the black solid line, the contributions from dimension-9 and dimension-5 operators are equal. The orange dashed lines show the current exclusion limit from KamLAND-Zen ($3.8 \cdot 10^{26}$ years) and the yellow dashed lines represent the expected future sensitivity ($2 \cdot 10^{28}$ years).  }
    \label{fig:half_lives_Volkas_LECs}
\end{figure}

The calculations for $0\nu\beta\beta$ in Sec.~\ref{sec:0vbb-Volkas} and the half-lives shown in
Fig.~\ref{fig:half_lives_Volkas} is based on the conservative assumption that all possible unknown LECs are zero, and the half-life calculated under this assumption gives an upper bound. Here, we briefly examine how much smaller the $0\nu\beta\beta$ half-lives can become when allowing for unknown LECs. Setting unknown LECs to their order-of-magnitude estimates, we obtain the half-lives shown in Fig.~\ref{fig:half_lives_Volkas_LECs} for the same benchmark scenarios. We see that the half-lives can be lowered by up to three orders of magnitude. However, the contribution from dimension-5 operators still dominates over that of dimension-9 operators in all cases. Moreover, the benchmarks \emph{B1} and \emph{B2}, for which the dimension-9 half-lives could be pushed down within the range of the future sensitivity, are excluded by cLFV observations.

\section{2-loop Weinberg diagrams for the 2-particle models}
\label{sec:2loopWein}
\begin{figure}[h]
    \centering
    \begin{subfigure}[t]{0.49\textwidth}
         \includegraphics[width=\linewidth]{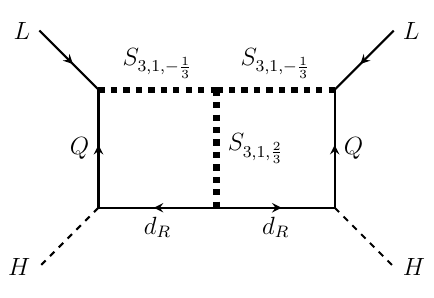}
         \caption{M5}
    \end{subfigure}
      \begin{subfigure}[t]{0.49\textwidth}
         \includegraphics[width=\linewidth]{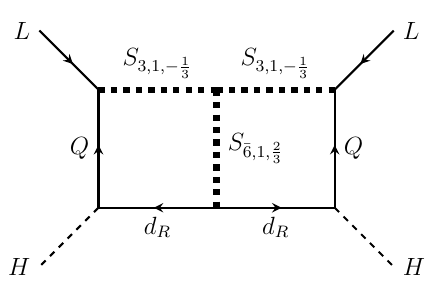}
         \caption{M6}
    \end{subfigure}
    \begin{subfigure}[t]{0.49\textwidth}
         \includegraphics[width=\linewidth]{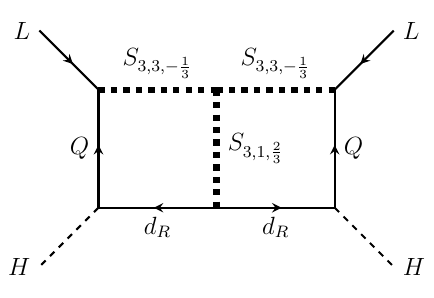}
         \caption{M7}
    \end{subfigure}
      \begin{subfigure}[t]{0.49\textwidth}
         \includegraphics[width=\linewidth]{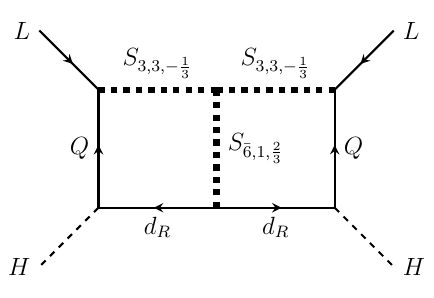}
         \caption{M8}
    \end{subfigure}
    \begin{subfigure}[t]{0.49\textwidth}
         \includegraphics[width=\linewidth]{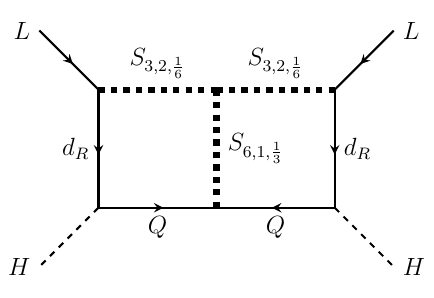}
         \caption{M9}
    \end{subfigure}
      \begin{subfigure}[t]{0.49\textwidth}
         \includegraphics[width=\linewidth]{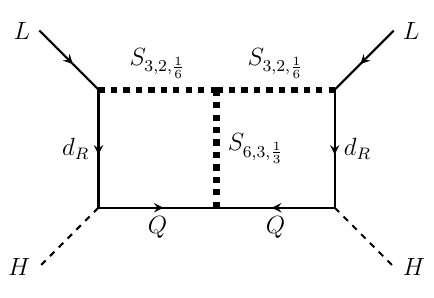}
         \caption{M10}
    \end{subfigure}
    \caption{CLBZ-1 based 2-particle 2-loop Weinberg models}
    \label{fig:CLBZ1}
\end{figure}

\begin{figure}[t]
    \centering
    \begin{subfigure}[t]{0.49\textwidth}
         \includegraphics[width=\linewidth]{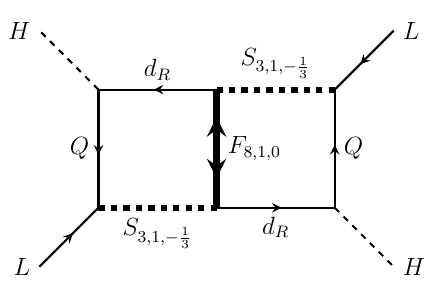}
         \caption{M11}
    \end{subfigure}
      \begin{subfigure}[t]{0.49\textwidth}
         \includegraphics[width=\linewidth]{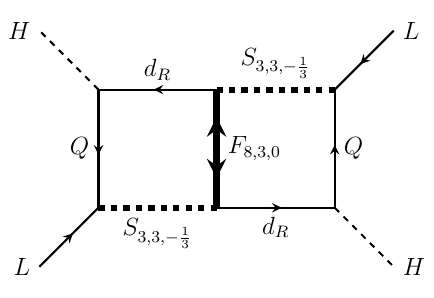}
         \caption{M12}
    \end{subfigure}
   \begin{subfigure}[t]{0.49\textwidth}
         \includegraphics[width=\linewidth]{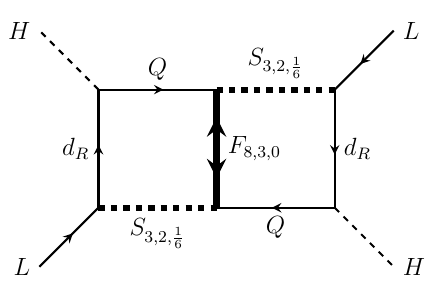}
         \caption{M13}
    \end{subfigure}
     \begin{subfigure}[t]{0.49\textwidth}
         \includegraphics[width=\linewidth]{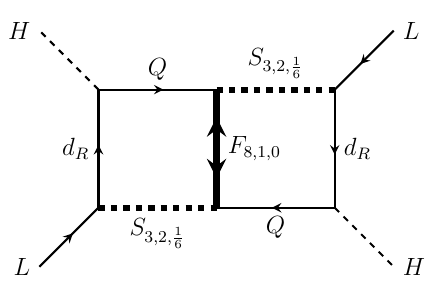}
         \caption{M14}
    \end{subfigure}
    \caption{PTBM-1 based 2-particle 2-loop Weinberg models}
    \label{fig:PTBM1}
\end{figure}

Here, we give the 2-loop diagrams for the Weinberg operator generated by the minimal 2-particle models in Tab.~\ref{tab:classification_minimal}. Models with two BSM scalars lead to the diagram type CLBZ-1, cf.~Fig.~\ref{fig:CLBZ1} and models with one BSM scalar and one BSM fermion to the diagram type PTBM-1, cf.~Fig.~\ref{fig:PTBM1}. Note that M5, M7 and M9 will not produce diagonal entries in the neutrino mass matrix, see Sec.~\ref{sec:2L_classification} for details.

\clearpage
\bibliographystyle{JHEP}
\bibliography{references}

\end{document}